\documentclass[amsmath,amsfonts,preprint,onecolumn,showkeys,floatfix,superscriptaddress,nofootinbib]{revtex4}

\usepackage{graphicx}
\usepackage{amssymb}
\usepackage{amsmath}
\usepackage{epstopdf}
\usepackage{float}
\usepackage{array}

\begin{document}

\title{Stable Frank-Kasper phases of self-assembled, soft matter spheres}

\author{Abhiram Reddy} 
\affiliation{Department of Polymer Science and Engineering, University of Massachusetts, Amherst, MA 01003}
\author{Michael B. Buckley} 
\affiliation{Department of Physics, University of Massachusetts, Amherst, MA 01003}
\author{Akash Arora} 
\affiliation{Department of Chemical Engineering and Materials Science, University of Minnesota, Minneapolis, MN 55455}
\author{Frank S. Bates} 
\affiliation{Department of Chemical Engineering and Materials Science, University of Minnesota, Minneapolis, MN 55455}
\author{Kevin D. Dorfman}
 \affiliation{Department of Chemical Engineering and Materials Science, University of Minnesota, Minneapolis, MN 55455}
\author{Gregory M. Grason} 
\affiliation{Department of Polymer Science and Engineering, University of Massachusetts, Amherst, MA 01003}

\begin{abstract} 

Single molecular species can self-assemble into 
 Frank Kasper (FK) phases, finite approximants of dodecagonal quasicrystals,
defying intuitive notions that thermodynamic ground states are maximally symmetric.  
FK phases are speculated to  emerge as the minimal-distortional packings of space-filling spherical domains, but a precise 
quantitation of this distortion 
and how it affects assembly thermodynamics 
remains ambiguous.  We use two complementary approaches to demonstrate that the principles driving FK lattice formation in diblock copolymers emerge directly from the strong-stretching theory of spherical domains, in which minimal inter-block area competes with minimal stretching of space-filling chains.
The relative stability of FK lattices is studied first using a diblock foam model with unconstrained particle volumes and shapes, which correctly predicts not only the equilibrium $\sigma$ lattice, but also the unequal 
volumes of the  equilibrium domains.  We then provide a molecular interpretation for these results via  self-consistent field theory, 
illuminating how molecular stiffness regulates the coupling between intra-domain chain configurations and the asymmetry of local packing. 
These findings shed new light on the role of volume exchange on the formation of distinct FK phases in copolymers, and suggest a paradigm for formation of FK phases in soft matter systems 
in which unequal domain volumes are selected by the thermodynamic competition between distinct measures of shape asymmetry.
\end{abstract}


\maketitle

\section{Introduction}

Spherical assemblies occur in nearly every class of supramolecular soft matter, from lyotropic liquid crystals and surfactants, to amphiphillic copolymers~\cite{Hyde97}.  In concentrated or neat systems, self-assembled spherical domains 
behave as giant ``mesoatoms,'' adopting  periodically-ordered crystalline arrangements.  While superficially similar to lattices formed in atomic or colloidal systems --  which are stabilized largely by bonding or translational entropy -- the periodic order in soft materials is governed by distinctly different principles because lattice formation occurs in thermodynamic equilibrium with the formation of the ``mesoatoms'' from the constituent molecules themselves.  Thus, the equilibrium sizes and shapes of  ``mesoatoms'' are inextricably coupled to the lattice symmetry, and vice versa.  

In this article, we address the emergence of non-canonical, Frank Kasper (FK) lattices in soft materials, characterized by complex and large-unit cells 
yet 
formed by assembly of a single molecular component.  Initially constructed as models of metallic alloys~\cite{FrankKasper59, Shoemaker86}, FK lattices are a family of periodic packings~\cite{Nelson89, Sadoc} whose sites are tetrahedrally-close packed and 
can be decomposed into polyhedral (e.g. Voronoi or Wigner Seitz) cells surrounding each site containing 12, 14, 15 or 16 faces. Known as the FK polyhedra, these cells (Z12, Z14, Z15 and Z16) possess variable volume and  in-radii. Hence, FK lattices are natural candidates to describe ordered, locally-dense packings of spherical elements of different radii such as atomic alloys~\cite{Shoemaker86, Sadoc} or binary nanoparticle superlattices~\cite{Travesset17}.  Once considered anomalous in soft matter systems, the past decade has seen an explosion in the observation of FK lattices in a diverse range of sphere-forming assemblies.  These include (A15, $\sigma$) liquid-crystalline dendrimers ~\cite{Balagurusamy97, Zeng04}, linear  ($\sigma$,A15) tetrablock~\cite{Lee10,Chanupriya16}, ($\sigma$, A15, C14, C15) diblock~\cite{Lee14, Kim17_1, Schulze17}  and  (A15) linear-dendron~\cite{Wiesner04} block copolymer melts, (A15) amphiphilic nanotetrahedra~\cite{Huang15,Cheng16}, (A15, 
$\sigma$, C14, C15) concentrated ionic surfactants~\cite{Mahanthappa17, Baez18} and (C14) monodisperse, functionalized nanoparticles~\cite{Pansu15}.  The central puzzle surrounding the formation of FK lattices in these diverse systems is understanding why single-components assemble into phases  composed of highly heterogeneous molecular environments.  

\begin{figure}
\center \includegraphics[width=0.45\textwidth]{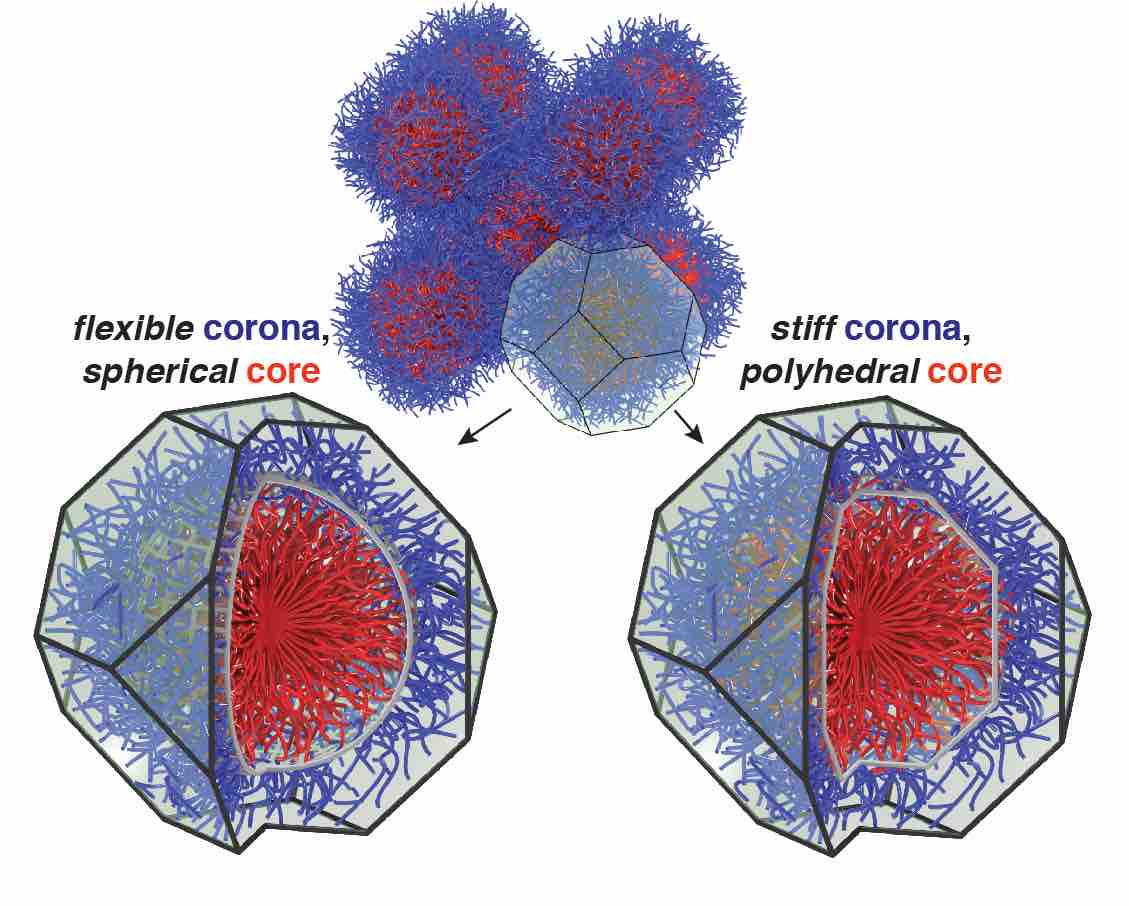}
\caption{\label{fig:1} Chain packing of spherical diblock copolymer domains of the BCC lattice (top), with corresponding limits of weakly-coupled (bottom left) and strongly-coupled (bottom right) of core domain shape of polyhedral (truncated-octohedron) cell symmetry.}
\end{figure}

A common element distinct to FK formation in soft systems is the thermodynamic cost of {\it asymmetry} 
imposed by incompatibility between uniform density and packing of perfectly spherical objects (Fig.~\ref{fig:1}).   In soft assemblies, the ideally spherically symmetric  domains are warped into lower-symmetry, polyhedral shapes (i.e. topologically equivalent to the Voronoi cells) which fill space without gaps. The minimal free energy state is the one for which the quasi-spherical domains (qSD) remain ``most spherical.''  The most commonly invoked notion of {\it sphericity} in this context is the dimensionless cell area $A$ to volume $V$ ratio, ${\cal A} \equiv A/( 36 \pi V^2)^{1/3}$, which has a lower bound of 1 achieved by perfect spheres.  The cellular partitions of FK lattices play a key role in the mathematical modeling of dry foams, known as the {\it Kelvin problem}~\cite{Weaire94, Phelan96, Kusner96, Cox17}, which seeks minimal area of partitions of space into equal volume cells.~\footnote{The tetrahedral coordination of FK lattice implies that their partitions closely approximate the geometric constraints of Plateau borders, and are therefore near to minimal-area partitions.  In addition to A15, at least two more partitions of FK lattices, $\sigma$ and H, have also been shown previously~\cite{Phelan96, Cox17} to beat the area of optimum originally conjectured by Kelvin, the BCC partition.  In Appendix \ref{Kelvin}, we report that the FK lattice P also belongs to this rarified category.}   Based on the fact that the lowest-area, equal-volume cellular partition known to date, the Weaire-Phelan foam~\cite{Weaire94}, derives from the FK lattice A15, Ziherl and Kamien proposed that this lattice is generically favored thermodynamically in so-called ``fuzzy colloid'' models~\cite{Ziherl00, Ziherl01}, an argument subsequently adapted to sphere phases of block copolymers~\cite{Grason03, Grason06}.  Recently, Lee, Leighton and Bates reasoned that average ``sphericity'' could be increased (i.e. decreased mean ${\cal A}$) below the Weaire-Phelan structure if the equal-volume constraint for distinct cells is relaxed, as would occur for molecular exchange between distinct qSD~\cite{Lee14}.  Based on the Voronoi partitions, which have unequal volumes for FK lattices, $\sigma$ was argued to have lower mean dimensionless area than A15, and thus should be stable over that lattice according to the sphericity argument, consistent with observations of  a $\sigma$ lattice in diblock copolymer melts~\cite{Schulze17} and self-consistent field theory of conformationally and architecturally asymmetric diblocks~\cite{ACShi14}. 

While the role of volume asymmetry has been implicated previously in the formation FK lattices by soft qSD assemblies~\cite{Glotzer11}, critical questions remain unanswered. First, for a given lattice, precisely which cell geometries and volumes accurately model qSD formation?   Second, what are the relevant measures of sphericity selected by the assembly thermodynamics? Finally, how do these in concert select the optimal balance between shape asymmetry (non-spherical domains) and volume asymmetry (molecular partitioning among domains) for a given qSD lattice, and in turn, select the equilibrium lattice and determine the scale of thermodynamic separation between the many competing FK lattices?  We address these questions in the context of what we call the {\it diblock foam model} (DFM), which quantifies the thermodynamic cost of asphericity in terms of a geometric mean of reduced cell area and dimensionless radius of gyration of the cells, and thus, integrates elements of both the {\it Kelvin} and {\it lattice Quantizer} problems~\cite{Conway88}.  These geometric proxies for inter-block repulsion and intra-molecular stretching in qSD exhibit qualitatively different dependencies on cell shape, a factor that we show, based on this model and self-consistent field theory (SCFT) analysis, to be critical to the volume partitioning among distinct qSD and optimal lattice selection.

Amongst the various classes of FK-forming soft matter \cite{Balagurusamy97, Zeng04, Lee10, Lee14, Kim17_1, Schulze17, Wiesner04, Huang15, Mahanthappa17, Pansu15}, we posit that diblock copolymers represent the optimal starting point for investigating the selection of low symmetry FK phases by soft matter spheres. Diblock copolymers are a relatively simple chemical system,  consisting of two flexible chains bonded together at their ends, and there exist robust theoretical methods for studying their phase behavior in the context of universal physical models~\cite{Matsen02, Arora16}. The fundamental mechanisms underlying assembly of diblock copolymers that we elucidate here furnish the foundation for subsequent investigations of other soft matter systems, where these basic principles are conflated with additional phenomena emerging from electrostatics, hydrophobic interactions, and detailed packing of the complicated (non-Gaussian) configurations of their constituents.

\section{Diblock Foam Model of FK Lattice Selection}
\label{DFM}
We adopt what we call the {\it diblock foam model} (DFM), first developed by Milner and Olmsted, in which the free energy of competing arrangements is reduced to purely geometric measures of the cellular volumes enclosing the qSD \cite{Milner94, Milner98}.  To a first approximation, these cells are the polyhedral Voronoi cells for a given point packing, whose faces represent coronal brushes flattened by contact with neighboring qSD coronae.  The model is based on strong-stretching theory (SST) of diblock copolymer melts, 
in which inter-block repulsions drive separation  into sharply divided core and coronal domains and the chains are well-extended.  
We also consider the case of large elastic asymmetry between core and coronal blocks, which itself derives from asymmetry of the block architecture or the segment sizes.  This corresponds to the  polyhedral-interface limit~\cite{Grason05}, in which the core/coronal interface in each qSD adopts a perfect, affinely shrunk copy of the cell shape (see Fig.~\ref{fig:1},bottom right).  Polyhedral warping of the interface is favored when the stiffness of the coronal blocks, which favors a more uniform extension from the interface to the outer cell wall, dominates over entropic stiffness of core blocks and inter-block surface energy, which both favor round interfaces.

\begin{figure}
\center
\includegraphics[width=0.95\textwidth]{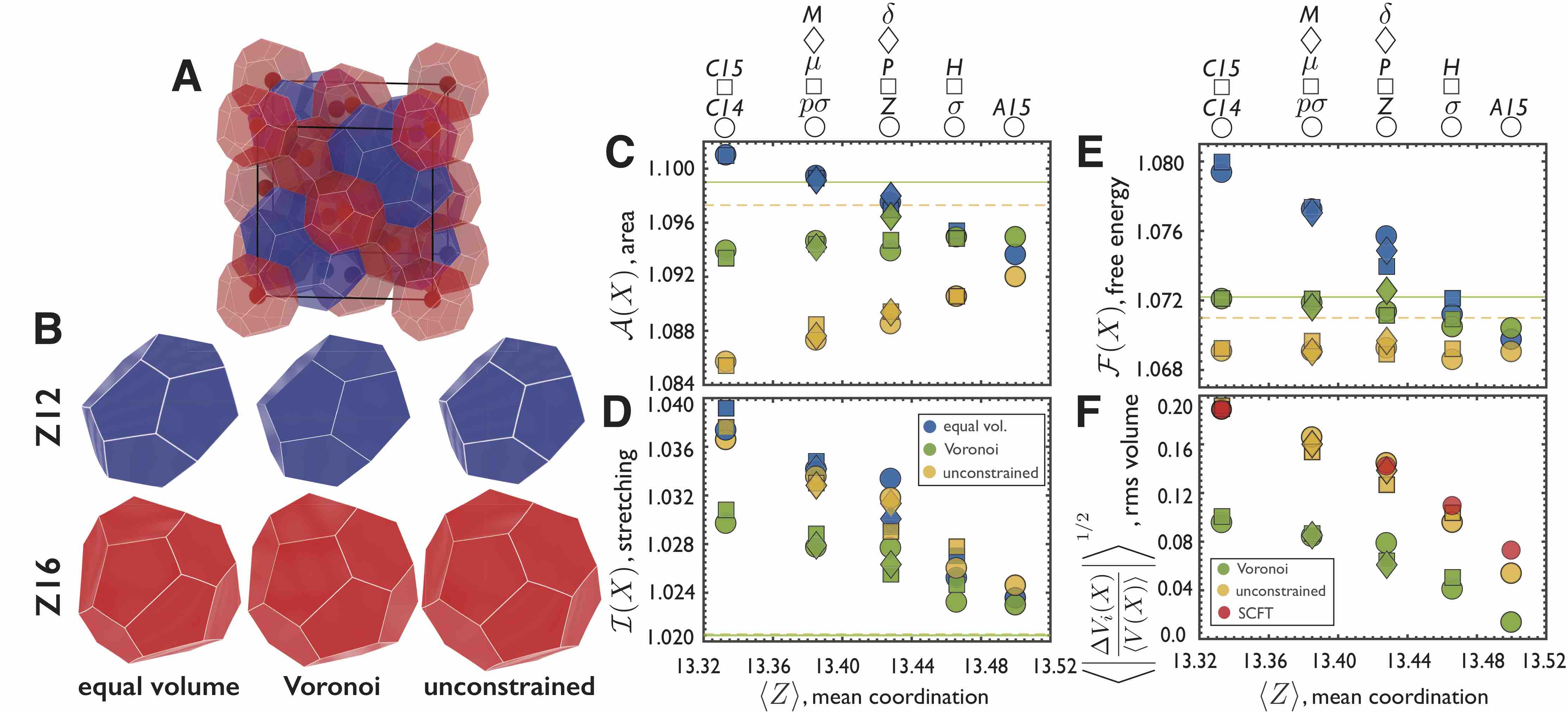} 

\caption{\label{fig:2} A DFM structure for the cubic repeat of C15 is shown in (A), qSD centers shown within the Z12 and Z16 cells, red and blue, respectively.  In (B), equilibrium shapes for three cell geometries studied, in which slight curvature of cell faces and edges  is visible for the relaxed shape cases.   Results of the DFM are shown for 11 competing FK phases (labeled above), plotted as function of mean coordination, or average number of cell faces $\langle Z \rangle$:   mean dimensionless area (C); mean dimensionless stretching (D); mean free energy (E); and rms volume variation among cells relative mean volume $\langle V(X) \rangle$ (F).  In (C)-(E) points are labeled according to the legend in (D) and the dashed and solid lines shows unconstrained and Voronoi results respectively for BCC.  In (F), variable volume cell results are compared to qSD volumes extracted from SCFT at $\chi N =40$, $f=0.25$ and $\epsilon=2$ as described in the  text.}
\end{figure}

In this limit, the free energy per chain~\cite{Milner98, Grason03}, $F(X)$, of a given lattice packing $X$ derives from two contributions,
\begin{equation}
\label{eq: 1}
F(X) = \gamma \frac{ {\cal A}(X)}{R_0} + \frac{\kappa}{2}~ {\cal I}(X) R_0^2 ,
\end{equation}
where $\gamma$ and $\kappa$ are coefficients fixed by the chain properties (i.e., block lengths, segment lengths, inter-block repulsion), and $R_0$ is the radius of a sphere of equal volume to the mean volume of cells, or $(4 \pi/3) R_0^3 = n_X^{-1} \sum_{i=1}^{n_X} V_i$, where $V_i$ is $i^{\rm th}$ cell volume of $n_X$ total cells in $X$ (see Appendix \ref{DFMdetails} for details).  The first term represents the enthalpy of core-corona contact, and hence is proportional to the (per volume) interfacial area, which itself is proportional to the cell area $A_i$, measured by the dimensionless (mean) cell area,  ${\cal A}(X) =(n_X^{-1} \sum_{i=1}^{n_X} A_i )/( 4 \pi R_0^2)$.  The second term represents the entropic costs of extending polymeric blocks (here modeled as Gaussian chains) in radial trajectories within qSD.  This cost grows with the square of domain size and depends on qSD shape through the dimensionless square radius of gyration, or  stretching moment ${\cal I}(X) =(n_X^{-1} \sum_{i=1}^{n_X} I_i )/( 4 \pi R_0^5/5)$ where $I_i = \int_{V_i} d^3 {\bf x}~|{\bf x} - {\bf x}_i|^2$ is the second-moment volume of the $i^{\rm th}$ cell, whose center lies at ${\bf x}_i$.  Optimizing mean cell size ($R_0$) yields the minimal free energy of lattice $X$, relative to the perfect sphere free energy $F_{0} =\frac{3}{2}(\gamma^2 \kappa)^{1/3} $,
\begin{equation}
{\cal F}(X) \equiv {\rm min}_{R_0} \big[F(X) \big]/F_{0}= \big[{\cal A}^2(X) {\cal I}(X)\big]^{1/3}.
\end{equation}
This geometric mean favors simultaneously low values of dimensionless area and stretching\footnote{Assembly thermodynamics depends on the dimensionless ratios of structure-averaged area and stretching of cells and volume, as opposed to averages of dimensionless cell area and stretching.}.   While minimal area partitions (at constant volume) are associated Kelvin's foam problem, lattice partitions that optimize $I_i$ (at fixed density) are the object of the {\it Quantizer problem}~\cite{Conway88}, which has applications in computer science and signal processing~\cite{Du99}.

The Milner and Olmsted model has been studied for flat-faced Voronoi cells of FCC, BCC and A15~\cite{Milner98, Grason06}, showing that the latter FK lattice of sphere-forming diblocks is favored over former two canonical packings in the polyhedral interface limit.  Here, we analyze a vastly expanded class of 11 FK lattices, possessing up to 56 qSD per periodic repeat.  Most critically, we employ a Surface Evolver~\cite{Brakke92} based approach that minimizes ${\cal F}(X) $ over arbitrary volumes {\it and} shapes of constituent cells in the DFM structure (see Appendix \ref{SEdetails} for detailed method and tabulated results).

To assess the importance of relaxing volume and shape, consider the three distinct ensembles of qSD cells, shown for C15 in Fig.~\ref{fig:2}A,B. We have computed results for {\it equal-volume, relaxed-shape} cells, which cannot exchange mass, and {\it centroidal Voronoi cells}, which have fixed flat-face shapes but unequal volumes. The former ensemble neglects the possibility of mass exchange between micelles, while the second optimizes stretching~\cite{Du99} but is suboptimal in terms of cell area\footnote{Centroidal Voronoi cells have generating points at the centers of volume of the cell, and hence, for a given $X$ minimize the mean-square distance of all points to their corresponding central point (see Appendix \ref{Quantizer} for additional details)}. Neither model is realistic but they provide useful points of comparison for the unconstrained, {\it relaxed-volume and shape} cell model, which strictly minimize ${\cal F}(X)$ for given $X$. Fig.~\ref{fig:2}C shows that allowing both volume and shape to relax leads to a complete inversion of the trend of ${\cal A}(X)$ with $\langle Z \rangle$. Importantly, there is also a near degeneracy for the free energy of FK structures in Fig.~\ref{fig:2}E, which all lie within 0.08\% in ${\cal F}(X)$ (as compared to the relatively large $\approx 1\%$ spread for equal-volume qSD). These results confirm the critical role of volume exchange among asymmetric qSD in the thermodynamics of lattice formation~\cite{Lee14, Kim17_1}. Among these nearly degenerate, fully unconstrained DFM structures, the $\sigma$ phase overtakes A15 (minimal for fixed, equal volume) as the minimal energy phase (with next lowest energy for P), consistent with its observation upon in annealling~\cite{Lee14, Schulze17} as well as recent SCFT studies of conformationally asymmetric diblocks~\cite{ACShi14}. Notably, however, in the {\it relaxed-volume and shape} DFM, $\sigma$ possesses neither the minimal area (C14), nor minimal stretching (BCC).  Rather, its predicted stability results from the optimal compromise between these competing measures of domain asphericity.   

\begin{figure}
\center
\includegraphics[width=0.95\textwidth]{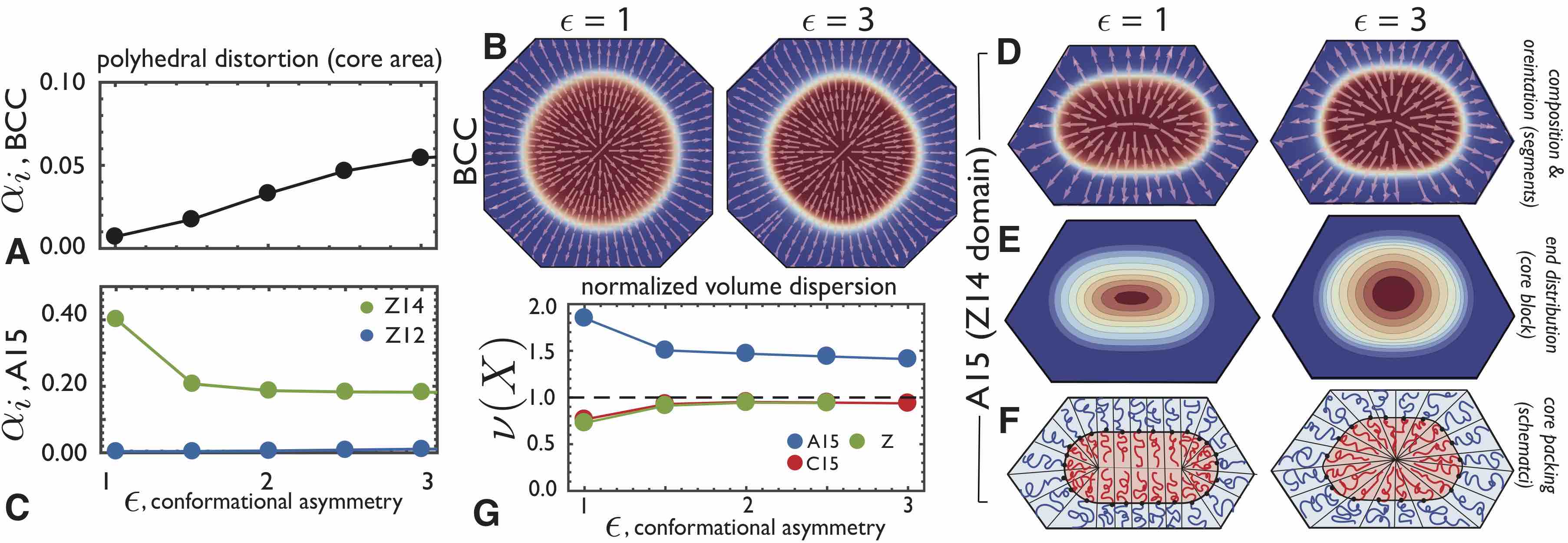} 
\caption{\label{fig:3} The polyhedral warping of the A/B interface, measured by $\alpha_i$ from SCFT profiles of $\chi N = 40$ and $f=0.29$ diblocks, of BCC qSD is plotted vs. conformational asymmetry $\epsilon = a_A/a_B$ in (A).  Corresponding 2D cross sections (normal to $[100]$ through center of primitive cell) of qSD within the truncated-octahedral cells of BCC are shown in (B), with composition varying from red in A-rich regions to blue in B-rich regions (A/B interface is white).  AAlso shown in vectors are the mean orientation of A-block
segments (polar order parameter)~\cite{Prasad17}.  In (C), the areal distortion of Z14 and Z12 qSD from SCFT predictions of A15 are shown (same composition and segregation strength as BCC), with corresponding section of the Z14 (cut normal to $[100]$ through face of primitive cell, see Fig. \ref{fig:sections}) qSD shown in (D) as in (B).  Additionally, spatial distribution of the A-block (core forming) chain ends are shown in (E), varying from zero density (blue) to maximal density (red) within the cores.   Schematics illustrating respective {\it discoidal} and {\it polyhedral} qSD packing are shown in (F). In (G), the volume dispersion (normalized by the DFM prediction) is plotted vs. conformational asymmetry.}
\end{figure}

The interplay between area and stretching underlies the emergent asymmetry in equilibrium qSD volumes.  Comparing the equal-volume to unconstrained DFM results in Fig.~\ref{fig:2}C and D shows that volume relaxation has a far more significant effect on relaxation of ${\cal A}(X)$ than ${\cal I}(X)$, which changes little by comparison.  Relaxation proceeds for all structures by inflating cells with relatively larger area, and shrinking smaller-area cells, restrained only by stretching cost creating highly unequal domain sizes (Fig.~\ref{fig:5}). Volume exchange for lattices with large proportions of lower area Z12 cells (e.g., C14 and C15) achieve relatively large ($\approx 2\%$) drops in ${\cal A}(X)$ when compared to the high-$\langle Z \rangle$ end of the spectrum (e.g., $\approx 0.2\%$ for A15).  

Cell volume asymmetry in equilibrated DFM structures pushes well beyond that of the ``natural'' geometry of Voronoi cells, which is strictly optimal for stretching, but not for its product with the square of dimensionless area.  Fig.~\ref{fig:2}F shows that both unconstrained and Voronoi models of qSD cell geometry exhibit an increase volume dispersion with decreasing mean coordination (or, with increasing fraction of Z12s). However, optimal unconstrained DFM cells are nearly twice as polydisperse in volume as the Voronoi distribution.  This massive volume asymmetry among qSD (up to $\approx 19\%$ variance for C14 and C15) is driven by dramatic reduction in inter-block contact area, a drive that is ultimately limited by the thermodynamic balance with the entropic (stretching) costs of filling space with qSD of unequal size.  These results imply that structures with a larger equilibrium volume dispersion (such as the lower-$\langle Z \rangle$ C14 and C15) structures are more susceptible to the effects of thermal processing that selectively promote or inhibit chain exchange among equilibrating spheres \cite{Kim17_1} than phases such as  A15, which relax free energy relatively little through volume equilibration.

Previous SCFT studies~\cite{ACShi14, Kim17_1} have shown that the canonical BCC sphere phase is overtaken by a stable $\sigma$ lattice when the elastic asymmetry, embodied by  ratios of statistical segment lengths $\epsilon \equiv a_A/a_B \gtrsim 1.5$.  DFM not only correctly predicts $\sigma$ as the dominantly stable sphere phase, but also does a remarkable job of predicting the relative hierarchy among metastable FK competitors.  This is evident in  Fig.~\ref{fig:scf_dfm}A-C, where we compare the free energies, scaled enthalpies and entropies for $\sigma$, Z, C14, C15 and A15 predicted by the unconstrained cell DFM to AB diblock SCFT calculations using methods described in ref.~\cite{Arora16} at somewhat strong segregation conditions $\chi N = 40$, where $\chi$ is the Flory-Huggins parameter for A/B contact and $N$ is the degree of polymerization.  DFM correctly predicts the narrow $~0.01\%$ scale of free-energy splitting between these competitors for $\epsilon = 2$ diblocks in the composition range $f\leq 0.25$, where $f$ is the volume fraction of the minority block.  Moreover, DFM predicts  their ranking relative to $\sigma$ with the exception of Z, which DFM predicts to be nearly degenerate with C15.   The accuracy of  DFM extends beyond thermodynamics to structure, most notably the volume asymmetry in Fig.~\ref{fig:2}F.

\section{Molecular Mechanism of Aspherical Domain Formation}

To probe the molecular mechanism that underlies the selection of FK lattices  in block copolymers, we analyze two order parameters that quantify the respective asymmetric shapes and volumes of qSD, computed from the volumes enclosing A-rich cores in SCFT composition profiles of diblocks at $\chi N = 40$,  $f=0.29$ and for variable conformational asymmetry (see Appendix \ref{GeometrySCFTdetails}).  The first parameter, \begin{equation}
\alpha_i = \frac{{\cal A}^{\rm A/B}_i - 1}{{\cal A}^{\rm poly} - 1},
\end{equation}
measures the degree of polyhedral warping of the core in terms of the dimensionless area ${\cal A}^{\rm A/B}_i$ of the A/B interface of the $i^{\rm th}$ domain relative to a sphere, where ${\cal A}_i^{\rm poly}$ is the dimensionless area predicted for the perfectly polyhedral interface of the corresponding cell from the unconstrained DFM:  $\alpha_i = 0$ for spherical interfaces; and $\alpha_i = 1$ for interfaces that adopt the polyhedral shapes of the DFM cells. We define a second parameter, $\nu(X)$, that measures asymmetry of unequal volumes enclosed within A/B interfaces predicted by SCFT, relative to the volume asymmetry predicted by polyhedral cells of DFM for the same structure $X$
\begin{equation}
\nu(X) =\frac{ \big\langle \big| \frac{ \Delta V_i(X)  }{ \langle V(X) \rangle } \big|^2 \big\rangle^{1/2}_{\rm A/B} } { \big\langle \big| \frac{ \Delta V_i(X)  }{ \langle V(X) \rangle } \big|^2 \big\rangle^{1/2}_{\rm poly} }
\end{equation}
where $\Delta V_i(X) = V_i - \langle V(X) \rangle$ is the volume deviation of the $i^{\rm th}$ domain relative to the average in $X$, and values of $\nu (X)$ greater (less) than 1 indicate that qSD in SCFT are more (less) polydisperse predicted by relaxed DFM cells.

It has been argued previously~\cite{Grason06} that the polyhedral warping, or faceting, of core-corona interfaces should increase with $\epsilon$, which controls the ratio of corona- to core-block stiffness, due to the relatively lower entropic cost of more uniformly stretched coronae achieved by polyhedral interfaces.  This expectation is consistent with the observed monotonic increase of $\alpha$ from 0 at $\epsilon =1$ to the saturated value of $\alpha \approx 0.05$ for $\epsilon \gtrsim 2-3$ for the qSD in BCC plotted in Fig.~\ref{fig:3}A\footnote{While this extends beyond what is realized with most flexible linear diblocks, bulky side chains including bottlebrush configurations and miktoarm polymers would make the upper limit accessible.}. As shown in Fig. \ref{fig:alpha}, the polyhedral warp of the interface grows also with increasing $f$, due to the increased proximity of the qSD cell boundary to the interface and relatively shorter coronal blocks at larger core fractions.  While clearly far from a sharply faceted shape, the increase in core shape anisotropy is obvious from 2D cuts through of the qSD shown in Fig.~\ref{fig:3}B, showing a visible warp of A/B interface towards the truncated-octahedral shape of the BCC cell at $\epsilon = 3$.

For the FK phases, which are composed of distinct-symmetry qSD, areal distortion exhibits a markedly different dependence on increased coronal/core stiffness, as illustrated by the plots of $\alpha_{12}$ and $\alpha_{14}$ vs. $\epsilon$ for A15 in Fig.~\ref{fig:3}C.  Z12 domains exhibit a monotonic, albeit modest, increase in distortion with $\epsilon$. Surprisingly, for the Z14 domains, the excess area drops from its maximal value of $\alpha_{14} \simeq 0.4$ in the conformationally symmetric case for $\epsilon =1$ down to a lower, yet significant plateau value of $\alpha_{14} \simeq 0.2$, roughly twice the areal distortion for BCC.  

The origin of this counterintuitive drop in dimensionless area of the Z14 cells with increased outer block stiffness is illustrated in Fig.~\ref{fig:3}D, which compares  2D sections of the Z14 qSD of A15 at $\epsilon =1$ and $\epsilon=3$.  While the shape for larger outer-block stiffness ($\epsilon = 3$) is consistent with a quasi-faceted interface that copies the polyhedral cell (with rounded edges) of the Z14 domain, the conformationally symmetric case ($\epsilon =1$) is neither faceted nor spherical. It instead adopts oblate, or {\it discoidal} shape.  The contrast in core shape is further reflected in the sub-interface (vector) orientational order parameter of A-segments~\cite{Prasad17} and the spatial distribution of A-block chain ends, also shown in Fig.~\ref{fig:3}D,E.  For larger $\epsilon$, the preference for more uniform coronal block stretching drives the quasi-polyhedral domain shape, with radial chain trajectories extending from the center of the domain, a point at which core block ends are concentrated.  In contrast, for the case of $\epsilon =1$, the stiffness of the core blocks is sufficient to resist deformations away from uniform core thickness.  Occupying the somewhat flattened Z14 cell with a qSD of uniform core thickness then leads to the discoidal shape, in which chain ends spread laterally in a quasi-lamellar core rimmed by a quasi-toroidal packing at its circumference. The preference for uniform core thickness within the relatively oblate Z14 cell, which gives rise to a larger area discoidal interface for $\epsilon =1$, ultimately gives way to the quasi-polyhedral qSD shape, and corresponding radial chain stretching, with increased outer block stiffness for $\epsilon \gtrsim 2$ (see schematic in Fig.~\ref{fig:3}F).

Fig. \ref{fig:sections} shows evidence of this same {\it discoidal} $\to$ {\it polyhedral} transition qSD within the most oblate cells of other FK phases, C15 and Z,  leading to a corresponding drop in excess area $\alpha_i$ from $\epsilon =1$ to $\epsilon \approx 2$ for those cells.  In Fig.~\ref{fig:3}G we find this intra-domain shape transition with increasing corona-/core-block stiffness is coupled to a transition in volume asymmetry among qSD.  Discoidal domains of the conformationally symmetric diblocks ($\epsilon=1$) realize a volume dispersion that is strongly divergent from the polyhedral geometry in the DFM, including both greater ($\nu(X) >1$, for A15) and lesser ($\nu(X) <1$ for Z, C15) dispersity. Yet, in the limit of $\epsilon \gtrsim 2$, relatively stiffer coronal blocks pull the cores into radial-stretching, quasi-polyhedral shapes.  This transition to more compact cores, in turn, results into volume redistributing among equilibrium qSD tending to the $\nu(X) \to 1$ limit, consistent with agreement between asymmetric volumes of DFM and SCFT shown in Fig.~\ref{fig:2}F.  

\begin{figure}
\center
\includegraphics[width=0.35\textwidth]{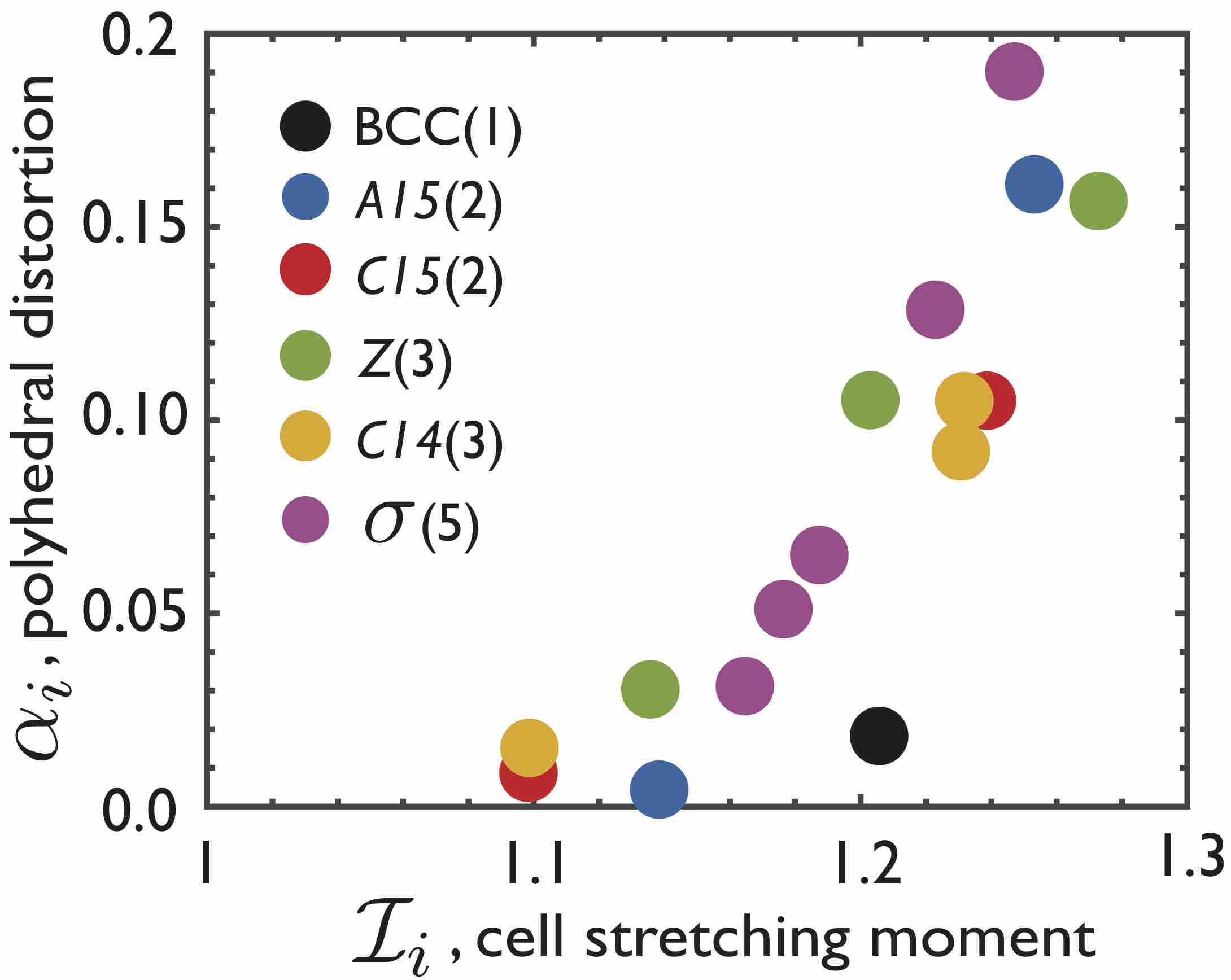} 
\caption{\label{fig:4} Correlation between polyhedral warping of core shapes ($\alpha_i$) within symmetry-distinct qSD extract from SCFT at $\chi N = 40$, $f=0.25$ and $\epsilon = 2$ and the degree of frustration of chain stretching in the correspond cell, quantified by the (cell-wise) dimensionless stretching moment, ${\cal I}_i$.}
\end{figure}

Notwithstanding the broad agreement between SCFT and DFM predictions, the degree of polyhedral warping of qSD shape is both arguably modest (i.e., $\alpha \lesssim 0.3$ for $\epsilon \gg 1$ for this $\chi N$ and $f$) and highly variable in the FK structures, suggesting a heterogeneous degree of shape frustration among cells.  Moreover, the {\it discoidal} $\to$ {\it polyhedral} transition occurs only in high-$\alpha$ qSD, whereas low-$\alpha$ cells (e.g., Z12 cells of A15) maintain radial stretching and a monotonic dependence on $\epsilon$.  What controls the variability of coupling between cell geometry of polyhedral distortion?  Fig.~\ref{fig:4} shows the correlation between $\alpha_i$ for qSD extracted from SCFT at $\chi N = 40$, $f=0.25$ and $\epsilon=2$ (i.e. in the quasi-polyhedral shape regime) plotted as a function the dimensionless stretching ${\cal I}_i$ for the corresponding cells from the DFM.  The generically increasing trend of $\alpha_i$ with ${\cal I}_i$ for cell geometries across competing phases argues that the variable degree of shape frustration within distinct qSD, and its consequent impact on qSD core shape, is regulated by the constraints of asymmetric chain-stretching in polyhedral cells.  In other words, the ultimate degree of asphericity of core distortion of qSD (measured by dimensionless area), is in fact, controlled by the local asphericity in radial stretching required by space-filling chain packing (measured by dimensionless radius of gyration). 

\section{Concluding Remarks}

We anticipate that the emergence of optimal FK lattice structure and thermodynamics via a balance of competing measures of domain asymmetry highlighted here for high molecular weight diblock copolymers will extend to other copolymer systems where these phases have been observed, including architecturally asymmetric copolymers, linear multiblocks, low molecular weight/high-$\chi$ systems and blends.  In particular, lower molecular weight polymers drive the system closer towards the strong segregation limit and away from the mean-field limit.
Each of these materials exhibit 
different molecular mechanisms through which the relative stiffness of the coronal domain transmits the asymmetry of the local qSD packing into the core shape.  For example, the observation of polygonal/polyhedral warping of outer zones of core-shell domains of linear mulitblock polymers~\cite{Gido93} provides a plausible mechanism to stabilize the $\sigma$ phase observed in linear tetrablocks~\cite{Lee10}.  On the other hand, accurately modeling the formation of $\sigma$ by low molecular weight conformationally asymmetric diblocks~\cite{Lee14, Schulze17}, likely requires a non-Gaussian (finite extensibility) model of chain stretching, but one which nevertheless, like the dimensionless radius of gyration ${\cal I}(X)$,  favors compact domains and competes against the minimal area preference for unequal domain volumes.  Beyond copolymers, we speculate further that additional intra- and inter-molecular mechanisms play the role of balancing the drive for minimal domain area in the formation FK phases, from giant nanotetrahedra~\cite{Huang15, Cheng16} to ionic surfactants~\cite{Mahanthappa17, Baez18}.

The present results for the DFM also shed new light on the non-equilibrium pathways for stabilizing metastable FK competitors, as has been demonstrated of conformational asymmetric linear diblocks quenched from high-temperature disorder sphere phases to low temperature metastable A15, C14 and C15 phases \cite{Kim17_1, Kim17_2}.  The low temperature quench is suspected to freeze out the inter-domain chain exchange needed to achieve the equilibrium $\sigma$ state, thus the kinetically-trapped quenched state inherits the volume distribution of high temperature micelle liquid state. The DFM suggests a new way to analyze the stability of FK states when domain volumes are out of equilibrium, suggesting the observation of C14 and C15 may be selected among the low-temperature kinetically trapped arrangements because it inherits a volume distribution that is both smaller in average cell size and possibly more polydisperse than the equilibrium state at the low temperature, and hence, a better fit to the ``aggregation fingerprint'' of low-$\langle Z \rangle$ packings.

\begin{acknowledgments}
R.Gabbrielli, J.-F. Sadoc, R. Mosseri and G. Schr\"oder-Turk are acknowledged for valuable input on geometric models of cellular packings. This research was supported by the AFOSR  under AOARD award  \#151OA107 and National Science Foundation under grants DMR-1719692 and DMR-1359191 (REU Site: B-SMaRT) .  G.G. also acknowledges the hospitality of the Aspen Center for Physics, supported by NSF PHY-1607611, where some of this work was completed. SCFT calculations were performed using computational facilities at the Massachusetts Green High Performance Computing Center and the Minnesota Supercomputing Institute.
\end{acknowledgments}


\appendix

\section{Diblock Foam Model} 

\subsection{Polyhedral Interface Limit of Strongly-Segregated Diblock Sphere Lattices}
\label{DFMdetails}
We briefly overview the strong-segregation theory (SST) calculation of Milner and Olmsted~\cite{Milner94,Milner98} for spherical domains in {\it polyhedral interface limit} (PIL), also known as the {\it straight path ansatz}.  We further show how the free energy of competing sphere packings is computed from purely geometric measures of the cellular volumes that enclose distinct spheres~\cite{Grason03,Grason06}, which forms the basis for the Diblock Foam Model (DFM).  

Here, we focus on the case of AB linear diblocks with conformational asymmetry, but the theory can be generalized to other architectures like miktoarm stars~\cite{Grason06}.  We consider a chain with total segment number $N=N_A+N_B$, with $f=N_A/N$ the fraction of the A-block. Segments are taken to have equal volumes $\rho_0^{-1}$ and potentially unequal statistical segment lengths, $a_{\rm A}$ and $a_{\rm B}$, for the respective blocks.  The ratio of segment lengths defines the conformational asymmetry $\epsilon \equiv a_{\rm A}/ a_{\rm B}$.  Within SST, the total free energy $F(X)$ (in units of $k_B T$) of a periodic repeat spherical assembly of lattice packing $X$  decomposes into two terms
\begin{equation}
F(X)=F_{int} + F_{st} ,
\end{equation}
which represents the respective costs of inter-block repulsions at a core/coronal interfaces and the entropic cost of stretching of (Gaussian) chains from random walk configurations.  

The first term $F_{int} = \Sigma A_{int}$ simply derives from the product of total area of core/coronal contact, $A_{int}$, times $\Sigma$ to give the surface area energy between phase separated A and B domains~\cite{Helfand75},
\begin{equation}
\Sigma= \rho_0 a \sqrt{\frac{\chi}{6}}  \Big( \frac{2}{3} \frac{\epsilon^{3/2}-\epsilon^{-3/2}}{\epsilon-\epsilon^{-1} } \Big) ,
\end{equation}
where $\chi$ is the Flory-Huggins parameter for AB repulsion and  $a \equiv \sqrt{a_{\rm A} a_{\rm B}}$ is the geometric mean of segment lengths\footnote{Note that $\Sigma$ is varies with conformational asymmetry $\epsilon$, and that it reduces to the standard result for interfacial tension between immiscible polymer melts in the symmetric limit $\Sigma(\epsilon \to 1) = \rho_0 a (\chi/6)^{1/2}$.}.  For the $i$th cell of $X$, the core/corona interface is an affinely shrunk copy of the outer cell that encloses a fraction $f$ of the total cell volume.  Hence, the area of the core interface of $i$th domain is $f^{2/3} A_i$, where $A_i$ is the cell area, and $A_{int} = f^{2/3} \sum_{i=1}^{n_X} A_i$, where $n_X$ is the number of domains (and cells) per periodic repeat.

The entropic contribution from chain stretching for domain $\alpha$ in cell $i$ (denoted as volume $V_{\alpha,i}$) can be evaluated using the SST entropy derived from the ``parabolic brush'' theory~\cite{Milner88}, which can be expressed as~\cite{Matsen02},
\begin{equation}
\label{eq: str}
F^{(\alpha)}_{st, i} = \frac{ 3 \pi^2 \rho_0}{8 N_\alpha^2 a_\alpha^2} \int_{V_{\alpha,i} } d^3{\bf x}~z^2 
\end{equation}
where $z$ is the distance from the AB interface, where junction points are localized, from which chain trajectories are assumed to extend along the radial lines extending from the cell center ${\bf x}_i$ to the outer wall of cell $i$, and $\int_{V_{\alpha,i} } d^3{\bf x}$ is the integral over volume. For spherical domains, these integrals are evaluated by describing the cell shape as a function of the radial directions $\hat{\Omega}$ extending from the cell center at ${\bf x}_i$:  $R_i (\hat{\Omega})$ and $R'_{i}(\hat{\Omega})$ are the respective distances to the interface and outer wall of the cell in direction $\hat{\Omega}$.  Because the core chains occupy a fixed fraction $f$ of each ``wedge'' in the PIL, we have $R'_{i}(\hat{\Omega}) = f^{1/2} R_i (\hat{\Omega})$, and the stretching contributions from each block are proportional to the same geometric stretching moment,
\begin{equation}
F^{(A)}_{st, i} = \frac{ 3 \pi^2 \rho_0}{8 N_A^2 a_A^2} \int d^2 \hat{\Omega} \int_0^{R'_{i}(\hat{\Omega}) } dr ~r^2 \big[R'_{i}(\hat{\Omega})  -r \big]^2 = \frac{ \pi^2 \rho_0 }{80 f^{1/3} N^2 a_A^2} S_i,
\end{equation}
and 
\begin{equation}
F^{(B)}_{st, i} = \frac{ 3 \pi^2 \rho_0}{8 N_B^2 a_B^2} \int d^2 \hat{\Omega} \int_{R'_{i}(\hat{\Omega}) }^{R_{i}(\hat{\Omega})}  dr ~r^2 \big[R'_{i}(\hat{\Omega})  -r \big]^2 = \frac{  \pi^2 \rho_0 (1-f^{1/3})^2(6+3f^{1/3}+f^{2/3}) }{80 (1-f)^2 N^2 a_B^2}  S_i ,
\end{equation}
where 
\begin{equation}
S_i \equiv \int d^2 \hat{\Omega} ~R^5_{i}(\hat{\Omega}) ,
\end{equation}
and is propotional to the second-moment of cell volume defined in the \ref{DFM},
\begin{equation}
I_i = \int_{V_i} d^3 {\bf x}~ |{\bf x} - {\bf x}_i|^2 =  \int d^2 \hat{\Omega} \int_0^{R_{i}(\hat{\Omega}) } dr ~r^4 = S_i /5 .
\end{equation}
Combining these together and summing over the cells in the periodic repeat we have the total stretching free energy
\begin{equation}
F_{st} = \frac{\pi^2 \rho_0}{16 N^2 a^2}\Big[\frac{f^{1/3}}{\epsilon} +\frac{  \epsilon (1-f^{1/3})^2(6+3f^{1/3}+f^{2/3}) }{(1-f)^2}\Big] \sum_{i=1}^{n_X} I_i .
\end{equation}

Since melt assembly occurs at fixed total density, equilibrium states correspond to states of minimal free energy per chain.  Defining the mean volume of the cells in $X$ as $V_0= n_X^{-1} \sum_{i=1}^{n_X} V_i$, the total number of chains per periodic repeat is $n_X V_0 \rho_0/N$.  The mean volume per cell also defines a measure of the mean cell dimension $R_0= (3 V_0/4 \pi)^{1/3}$, the radius of a sphere of equal volume to $V_0$.  Using this definition we can rewrite the area per volume as 
\begin{equation}
\frac{ \sum_{i=1}^{n_X} A_i}{n_X V_0} = \frac{3 {\cal A} (X)}{R_0}
\end{equation}
and the stretching per volume as
\begin{equation}
\frac{ \sum_{i=1}^{n_X} I_i}{n_X V_0} =\frac{3}{5} {\cal I} (X) R_0^2 
\end{equation}
where the dimensionless quantities ${\cal A} (X)=n_X^{-1} \sum_{i=1}^{n_X} A_i/(4 \pi R_0^2)$ and ${\cal I} (X)=n_X^{-1} \sum_{i=1}^{n_X} I_i/(4 \pi R_0^5/5)$ depend only on cell shapes and are independent of $R_0$, or mean domain size.   Using these quantities and dividing $F_{int}+F_{st}$ by the total chain number we arrive at \ref{eq: 1}, where the coefficients are given by,
\begin{equation}
\gamma = N a \sqrt{2 \chi/3}  \Big( \frac{\epsilon^{3/2}-\epsilon^{-3/2}}{\epsilon-\epsilon^{-1} } \Big),
\end{equation}
and
\begin{equation}
\kappa =  \frac{3\pi^2}{80 N a^2}\Big[\frac{f^{1/3}}{\epsilon} +\frac{  \epsilon (1-f^{1/3})^2(6+3f^{1/3}+f^{2/3}) }{(1-f)^2}\Big] ,
\end{equation}
which are independent of structure $X$ and are fixed for a given set chain properties.  Optimizing $F(X)$ with respect to $R_0$, we find a equilibrium mean domain size
\begin{equation}
\label{eq: R0}
(R_0)_{eq} = R_s \Big(\frac{ {\cal A}(X)}{ {\cal I}(X) } \Big)^{1/3} ,
\end{equation}
where $R_s = (\gamma/\kappa)^{1/3} \propto (\chi N)^{1/6} N^{1/2} a$ is the thermodynamically selected radius of domains if cells were equal volume spheres (i.e. ${\cal A}={\cal I}=1$).  

We note that this model relies on the so-called parabolic brush theory~\cite{Milner88} in the expressions for Gaussian chain entropy in eq. (\ref{eq: str}), which are known to fail for brush-like domains with convex curvature due to the presence end-exclusions zones missing from the parabolic model.  Notwithstanding, the failure to properly account for these exclusion zones in the coronal blocks of this calculation~\cite{Belyi04}, this approximation only modifies the coefficient $\kappa$ and its dependence of $f$\footnote{Specifically, it can be shown that the coronal brush free energy in a spherical geometry is proportional to $h^2$, where $h$ is the brush thickness, times a function of $h/R_s$, where $R_s$ is the spherical radius.  In this geometry $h(\hat{\Omega})=R(\hat{\Omega})-R'(\hat{\Omega})=(1-f^{1/3})R(\hat{\Omega})$ and $R_s=R'(\hat{\Omega})$ so that $h/R_s=(f^{-1/3}-1)$ for each wedge, independent of $\hat{\Omega}$}.  The proportionality of the stretching free energy with $\int d^2 \hat{\Omega} ~R^5(\hat{\Omega})$ follows on the general grounds that each ``wedge'' of the domain includes  a number of chains proportional to $d^2 \hat{\Omega}~R^3(\hat{\Omega})$, each of which is stretched a distance proportional to $R(\hat{\Omega})$ and hence acquires a free energy penalty proportional to $R^2(\hat{\Omega})$.

\subsection{Unequal Domain Volumes in A15 Lattices:  Weighted Voronoi Partitions}
\label{flat face}
Here we illustrate the dependence of the DFM energy ${\cal F} (X) = \big[{\cal A}^2 (X){\cal I} (X)\big]^{1/3}$ on the volume difference between symmetry-distinct cells of FK lattices.  For this purpose we consider the A15 lattice, which can be decomposed into two Z12 cells (at the center and corners of the primitive, cubic cell) and six Z14 cells (two positions decorating each face of the primitive, cubic cell).  An analytical relation for ${\cal F} ({\rm A15})$ can be obtained using expressions for cellular area, volume and second-moments of volume for the weighted Voronoi cells of A15.    Standard Voronoi partitions derive from the polyhedra constructed by planes that bisect the center-to-center neighbor separation vectors normally.   Here, we use the weighted partitions of A15 derived by Kusner and Sullivan~\cite{Kusner96}, correspond to the (flat-face) polyhedra constructed from planes at variable separation between the Z12 and Z14 sites (i.e., non-bisecting).

\begin{figure}
\center
\includegraphics[width=0.95\textwidth]{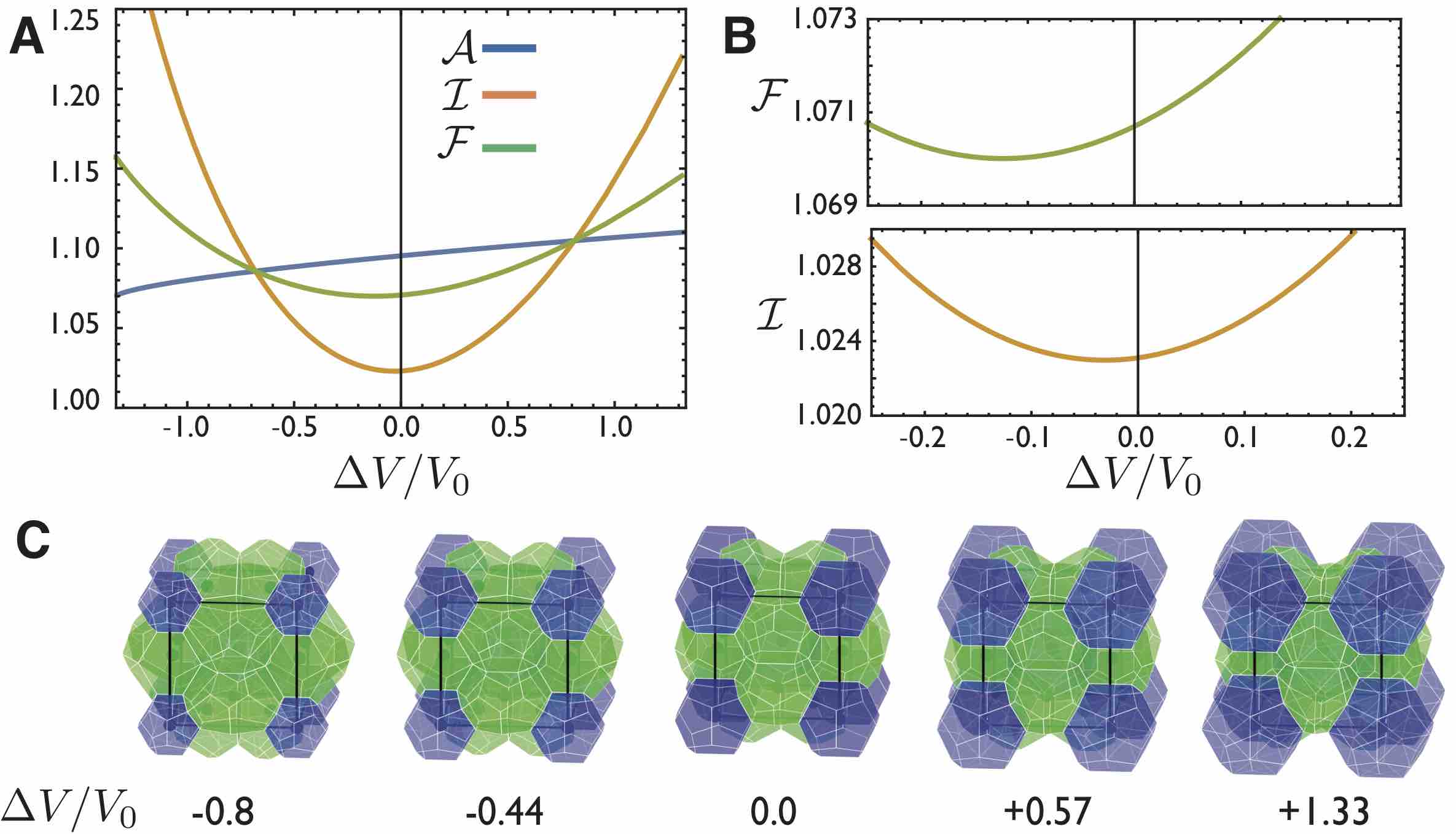}
\caption{\label{fig:5}  (A) Plot of $\mathcal{A}$,  $\mathcal{I}$ and  $\mathcal{F}$ for A15 lattice as a function of varying volumes of Z12 and Z14 cells; in (B) zoom in of minimal for free energy and stretching.  In (C), schematics of A15 primitive cell with unequal cell volumes: Z12 (blue) and Z14 (green) polyhedra.}
\end{figure}

Fixing the length of the primitive cubic cell to 2, and the mean volume per cell is fixed to $V_0=1$ (non-dimensional lengths), the total area of the cells can be expressed as a function of $c$, which parameterizes the size of the dodecahedral Z12 cells:  the volumes of these cells are $V_{\rm Z12} =  c^3/2$; which implies $V_{\rm Z14}=(4-V_{\rm Z12})/3$.  The mean cell area~\cite{Kusner96} is
\begin{equation}
\frac{1}{8}\sum_{i=1}^8 A_i(c) = \frac{3}{2} + \frac{3}{2} \sqrt{6} + \frac{1}{8}(6\sqrt{5} - 4\sqrt{6} - 3)c^{2}
\end{equation}
while the total second moments per cubic repeat was calculated by Kashyap and Neuhoff~\cite{Kashyap01} as
\begin{equation}
\frac{1}{8}\sum_{i=1}^8 I_i(c) =\frac{1}{32}(3c^{4}-5c^{3}+10) .
\end{equation}
Normalizing these by the area and second moment of spheres of $V_0=1$ gives the dimensionless free energy of A15 for flat faced cells,
\begin{equation}
{\cal F}_c ({\rm A15}) = \Big(\frac{5}{55296}\Big)^{1/3}  \Big(\big[12 + 12\sqrt{6} + (6\sqrt{5} - 4\sqrt{6} - 3)c^{2}]^2 (3c^{4}-5c^{3}+10)\Big)^{1/3}
\end{equation}
The dimensionless area, stretching and free energy are plotted as function of the volume difference 
\begin{equation}
\label{eq: delta}
\frac{\Delta V}{V_0} = \frac{2}{3} (c^3-2),
\end{equation}
where $\Delta V = V_{\rm Z12}-V_{\rm Z14}$ between Z12 and Z14 cells in Fig.~\ref{fig:5}.   This shows that the dimensionless area is minimal for vanishing Z12 volume ($c=0$), while dimensionless stretching is in fact minimized by the standard (centroidal) Voronoi partition ($c=5/4$), which has a very nearly equal volumes  $\Delta V/V_0 =-0.03$.  The competition between these drives for unequal cell volumes which results in optimal free energy (${\cal F}_c ({\rm A15}) = 1.070$) for flat faced cells with $c_{min}=1.22$ and $\Delta V/V_0 =-0.12$.

\section{Numerical optimization of cell geometry}
\label{SEdetails}
Beyond the case of the flat-faced A15 cell results shown in Fig.~\ref{fig:5} and discussed in the previous section, the relaxation of the cellular partitions from competing SD phases is performed using the {\it Surface Evolver} (SE)~\cite{Brakke92}.  While most commonly used for area optimization problems (e.g. dry foam models, minimal surfaces), SE generically optimizes a target function (e.g., energy, area) defined on the facets of triangulated surface mesh subject to various geometric constraints, for example periodic boundary conditions or volumes within bodies enclosed bodies (e.g., cells or bubbles).  

For the DFM (and for Kelvin and Quantizer problem results below) we construct initial configurations that are input into SE by generating the Voronoi partitions from the point lattice positions of competitor structures within triply periodic, rectilinear box.  The aspect ratio of the periodic cell dimensions and the initial coordinates of the cell centers are extracted from references listed for each FK lattice in Sec. \ref{results} below.  Voronoi cells are computed using $\it Voro++$ ~ \cite{Rycroft09} and then converted to SE input files via a custom python script.  In addition to the initial topology of the ``foam" structure, the SE input file also defines a discrete approximation of the dimensionless stretching, ${\cal I}$, area ${\cal A}$, and DFM free energy, ${\cal F}$.  

Dimensionless area derives directly from computed total facet area and enclosed volumes of cells, while the stretching is computed as follows.  For cell $i$ in structure $X$, an initial center ${\bf x}_i'$ is chosen as fixed reference point within the body.  Since  ${\bf x}_i'$ is, in general, not the centroid of $i$, the stretching integral $\mathbf{x}_i$ splits into two contributions,
\begin{equation}
I_i= I'_i({\bf x}_i') - V_i|{\bf x}_i-{\bf x}'_i|^2
\end{equation}
where $ I'_i({\bf x}_i') \equiv \int_{V_i} d^3 {\bf x} |{\bf x}-{\bf x}'_i|^2$ is the second moment of $V_i$ with respect to the reference point ${\bf x}'_i$.  For a triangulated mesh composed of triangular facets with center ${\bf X}_f$, area $\Delta A_f$, (outward) normal ${\bf N}_f$, these quantities can be approximated using the discrete sums,
\begin{equation}
I'_i({\bf x}_i')= \frac{1}{5} \sum_{f\in i} \Delta A_f {\bf N}_f \cdot ({\bf X}_f-{\bf x}'_i) | {\bf X}_f-{\bf x}'_i|^2 ,
\end{equation}
and 
\begin{equation}
{\bf x}_i-{\bf x}'_i= \frac{1}{4 V_i} \sum_{f\in i} \Delta A_f\big[ {\bf N}_f \cdot ({\bf X}_f-{\bf x}'_i) \big] ({\bf X}_f-{\bf x}'_i ) .
\end{equation}
These quantities are evaluated by use of {\texttt{ facet\_general\_integral}} in SE.  In the limit of $\Delta A_f \to 0$ these sums converge to the integral quantity with an error inversely proportional to the number of facets. 

Cell shape optimizations are performed in order to minimize dimensionless free energy (${\cal F}$), area  (${\cal A}$) or stretching (${\cal I}$) for a given set of constraints, fixed periodicity, number of cells and with or without enforcing equal volume among distinct cells.  Numerical optimization proceeds by successive interaction of vertex relaxation followed by mesh refinement steps.  For each mesh refinement, vertices are relaxed until the optimized quantity (${\cal F}$, ${\cal A}$ or ${\cal I}$) changes by less than $10^{-6}$.  Mesh refinements proceed until the total change of post-relaxation value of the target quantity falls below $10^{-6}$ upon successive mesh refinements.

\begin{figure}[H]
\center
\includegraphics[width=0.90\textwidth]{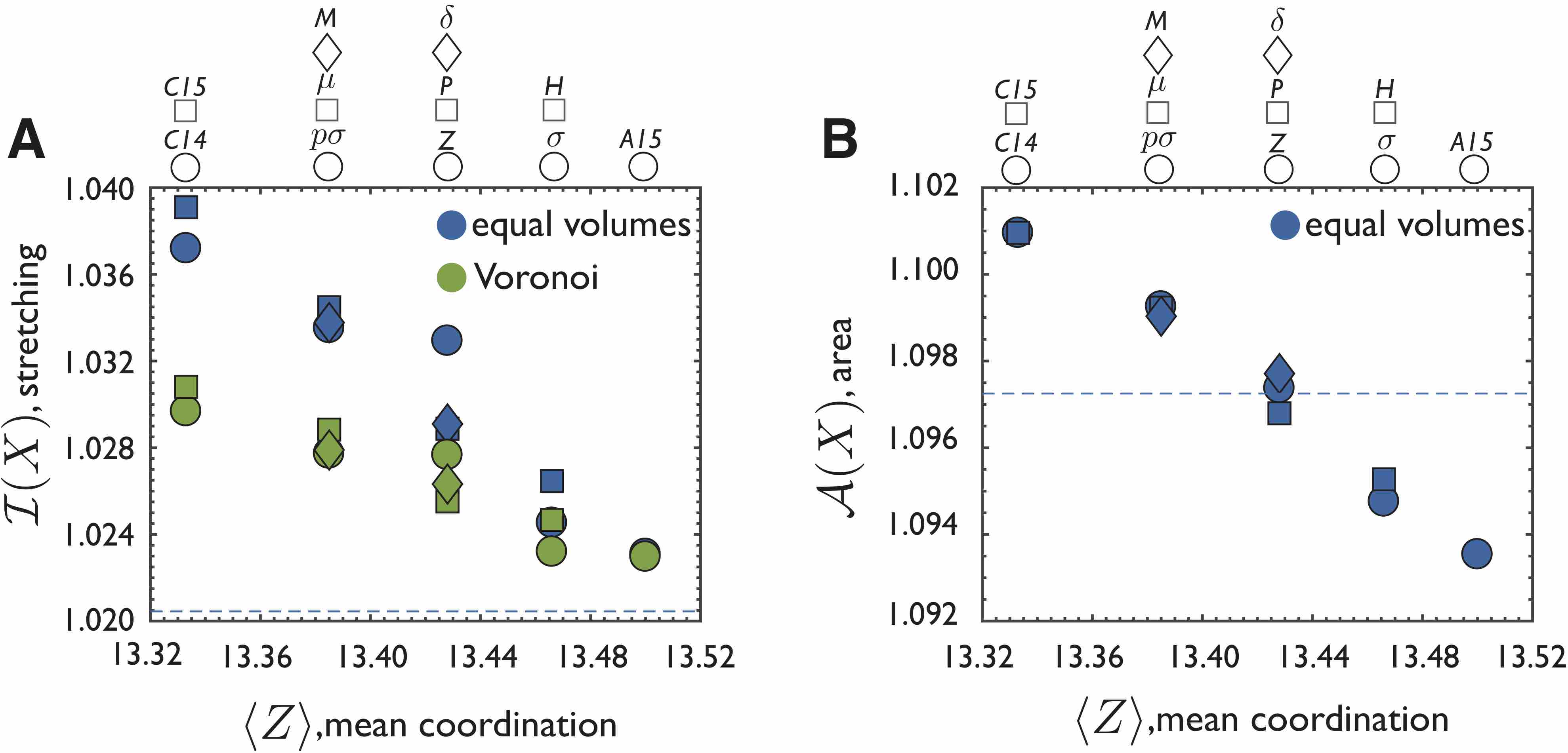}
\caption{\label{fig:6} (A) Computed minimal dimensionless stretching moments (${\cal I}(X)$) computed from:  equal volume constraint (blue) and Centroidal Voronoi Partitions (green). (B) Computed minimal dimensionless area (${\cal A}(X)$) for equal volume cells (i.e. Kelvin problem results). Dashed lines show results for BCC.}
\end{figure}

\subsection{Minimal stretching cells:  Quantizer problem}
\label{Quantizer}

For comparison to DFM result, we compute the cell geometries that optimize ${\cal I}$ for the competing FK lattices.  This optimization is directly related to the {\it Quantizer problem}~\cite{Kashyap01, Lloyd82, Barnes83} that seeks the optimal decomposition of lattice into cells, at a given cell density, whose average square distance to the generating points ${\bf x}_i'$ is minimal (i.e., minimal sum of the second moments $\int_{V_i} d^3{\bf x} |{\bf x}-{\bf x}_i'|^2$) ). For a fixed set of {\it generators}, ${\bf x}_i'$, the cells that minimize ${\cal I}$ are given by the (unweighted) Voronoi partition (VP) derived from ${\bf x}_i'$. Additionally, for a fixed set of {\it cells}, $V_i$, the generators that minimize ${\cal I}$ are given by the centroids ${\bf x}_i= V_i^{-1} \int d^3{\bf x} ~{\bf x}$ of $V_i$.  Hence, VP whose generating points are cell centroids, so-called {\it centroidal Voronoi partitions} (CVP), are local minima of ${\cal I}$, for a given cell topology~\cite{Du99, Du05}. Therefore, in the context of the DFM, the CVP correspond to cell geometries that rigorously minimize the entropic cost of chain stretching.

We compute the CVP for competing lattices by minimizing ${\cal I}(X)$ within SE starting with generating points corresponding to reported lattice site positions of FK lattices (as summarized in Sec. \ref{results}). CVP results from the minimization of  ${\cal I}$ at fixed mean volume of cells $V_0$ (fixed dimensions of the periodic repeat) but without constraints on the individual cell volumes $V_i$ are shown in Fig.\ref{fig:6}A.  For comparison, we also compute the minimal stretching cells for {\it fixed, equal cell volumes}, $V_i =V_0$, a constraint which accounts for increased values of ${\cal I}(X)$ that decrease with mean coordination of cells $\langle Z \rangle$.  This trend is consistent with Fig.\ref{fig:2} where the asymmetry in volume among cells for CVP is smallest for large-$\langle Z \rangle$ structures (A15) and largest at for small-$\langle Z \rangle$ structures (C14, C15).  This trend implies that imposing the equal cell volume constraints requires a smaller distortion from the optimal stretching cell geometry for structures whose CVP are closest to equal volume (i.e. at large-$\langle Z \rangle$ ).  Notably, no FK structure beats the minimal stretching value of BCC partition, ${\cal I}({\rm BCC})=1.0205$, proven to be the best quantizer among lattices in 3D~\cite{Barnes83}.

\subsection{Minimal area cells:  Kelvin problem}
\label{Kelvin}
For comparison to DFM result, we also compute the cell geometries that optimize ${\cal A}$ for the competing FK lattices.  As shown earlier in Sec.\ref{flat face} for flat faced cells (weighted VP) of A15, optimizing ${\cal A}$ in the absence of the volume constraints is unstable due to shrinking of cells to zero volume, and hence is not well-defined with respect to comparison of partitions of different lattices $X$.  A well-defined comparison is  possible with constraints on the relative volumes among cells, such as in the case of the {\it Kelvin problem}, which seeks partitions into {\it equal volume cells} ($V_i =V_0$) that minimize mean cell area, or equivalently, surface area energy.  The results from the minimization of ${\cal A}$ at fixed, equal volume of cells is shown in Fig. \ref{fig:6}B.  Notably, the minimal area partition deriving from A15, the Weaire Phelan foam, is minimal~\cite{Weaire94}, beating Kelvin's conjectured optimal foam, deriving from BCC.  Partitions of FK structures $\sigma$ and H also been previously reported ~\cite{Phelan96,Cox17} to achieve lower areas than the Kelvin foam\footnote{We note that reported values for dimensionless area by Phelan, which from Fig. 2 of  ref. \cite{Phelan96} gives ${\cal A}(\sigma) =1.094759>{\cal A}({\rm H})=1.095271$, are slightly discrepant (and smaller than) with our findings, reported in Sec. 2\ref{results}} (yet still larger than A15).  Here, we report a third counterexample of Kelvin's conjecture, deriving from the FK lattice P (${\cal A}({\rm BCC} ) = 1.097251>{\cal A}({\rm P} ) = 1.096795>{\cal A}({\rm A15} ) = 1.093541$) as well as minimal area (equal volume) results for four other previously unreported FK lattices, p$\sigma$, $\delta$ and M, all with dimensionless area exceeding the Kelvin foam.

\subsection{Competing lattices: Results}
\label{results}
Here, we summarize the SE results for competing structures in terms of dimensionless free energy $\mathcal{F}$, dimensionless surface area energy $\mathcal{A}$, dimensionless stretching moment $\mathcal{I}$ for various optimizations (optimal area, stretching or DFM free energy) and under different constraints (with and without equal cell volume constraints).  For comparison, we also include dimensionless area, stretching and DFM free energy for Voronoi partitions corresponding to the initial generating points (not necessarily centroidal).  Additionally, we compare volume histograms for both centroidal Voronoi partitions (upper histogram) and unconstrained DFM cells (lower histogram), plotted in terms of the volume fraction $\phi_i$ occupied by cell type $i$, in each structure.  For each structure, cell types are classified and color coded in terms of number of faces: Z12 (blue); Z14 (green); Z15 (orange); and Z16 (red).   Additionally, in the histograms, cell types are annotated according to the Wyckcoff positions corresponding to their generating points. 

\begin{figure}[H]
\center
\includegraphics[width=0.625\textwidth]{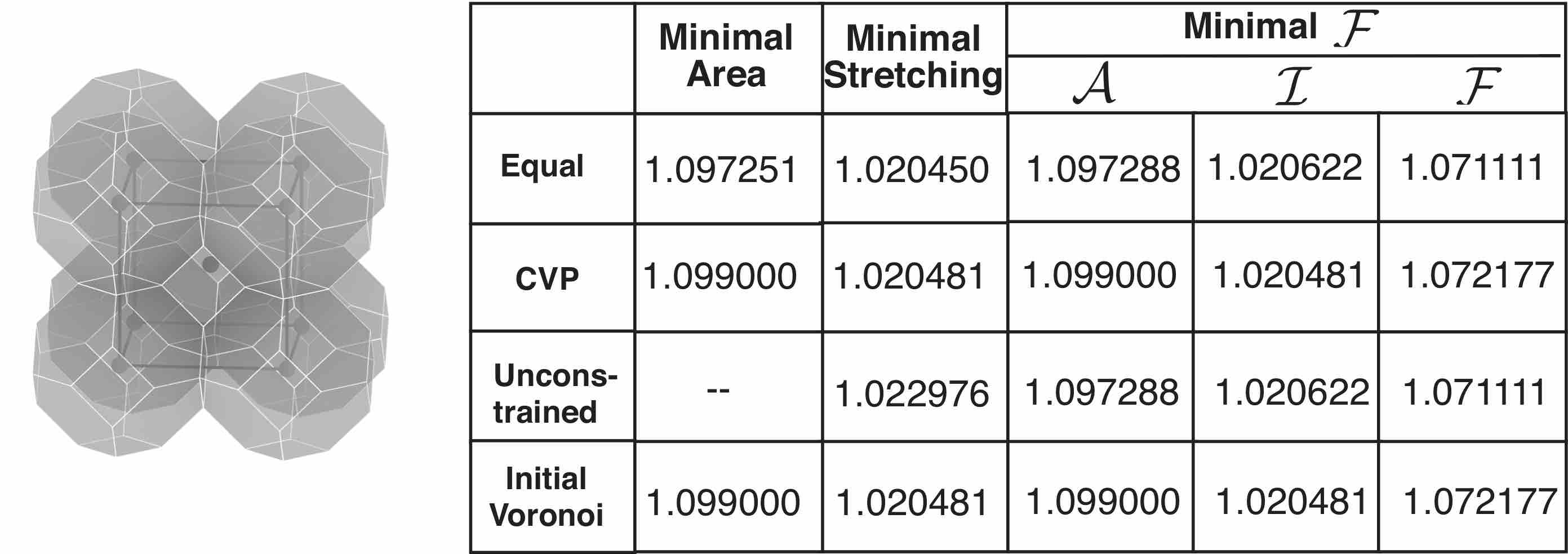}
\caption{BCC; space group: $Im\bar{3}m$; periodic cell:  (cubic) a:b:c=1:1:1; $n_{X}$ = 2, $\langle Z \rangle$ = 14}
\end{figure}

\begin{figure}[H]
\center
\includegraphics[width=0.7\textwidth]{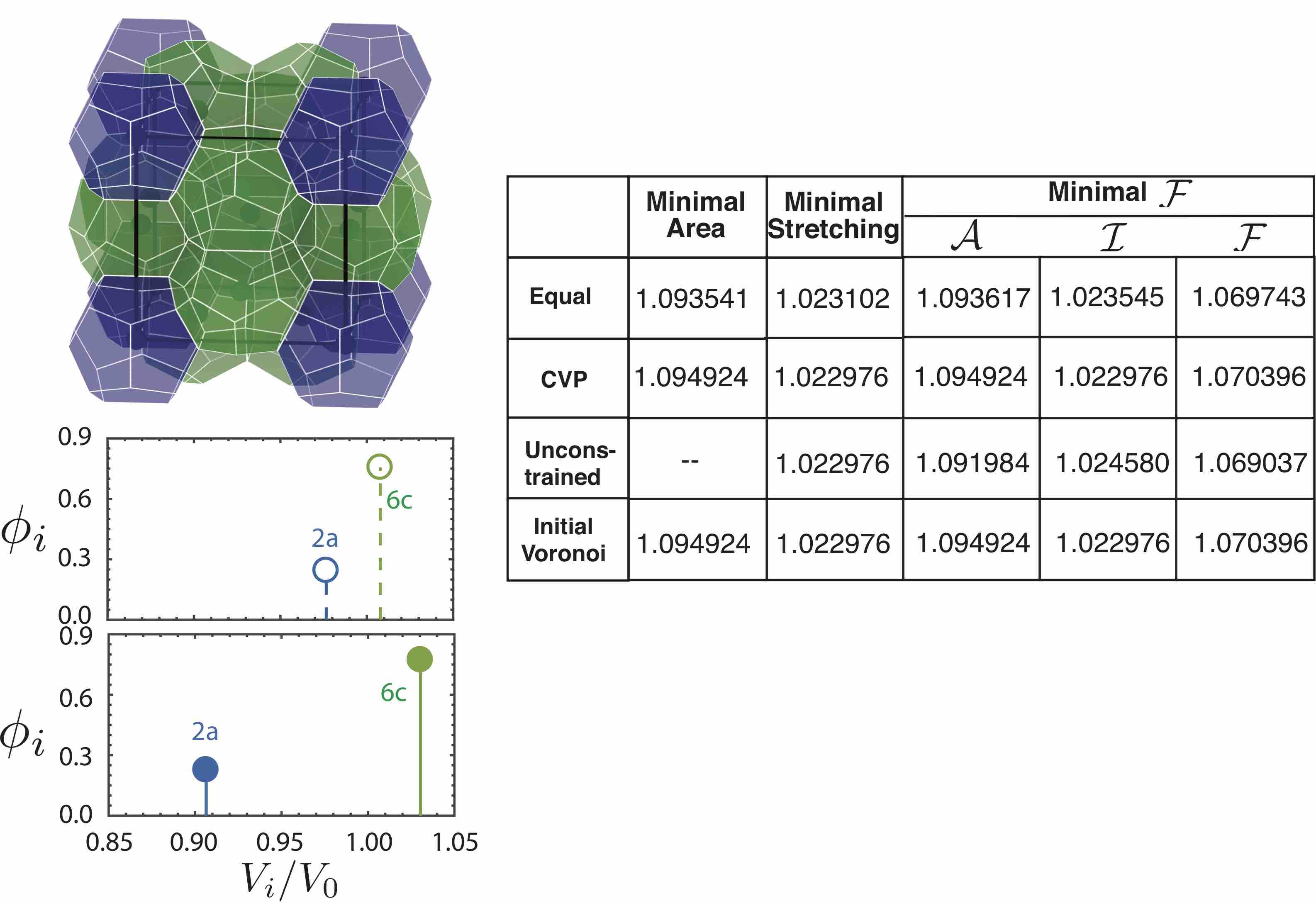}
\caption{$A15$; space group: $Pm\bar{3}n$; periodic cell:  (cubic) a:b:c=1:1:1; $n_{X}$ = 8; $\langle Z \rangle$ = 13.5, init. coords.: $Cr_{3}Si$ from ref. \cite{A15C14C15Mu}}
\end{figure}

\begin{figure}[H]
\center
\includegraphics[width=0.525\textwidth]{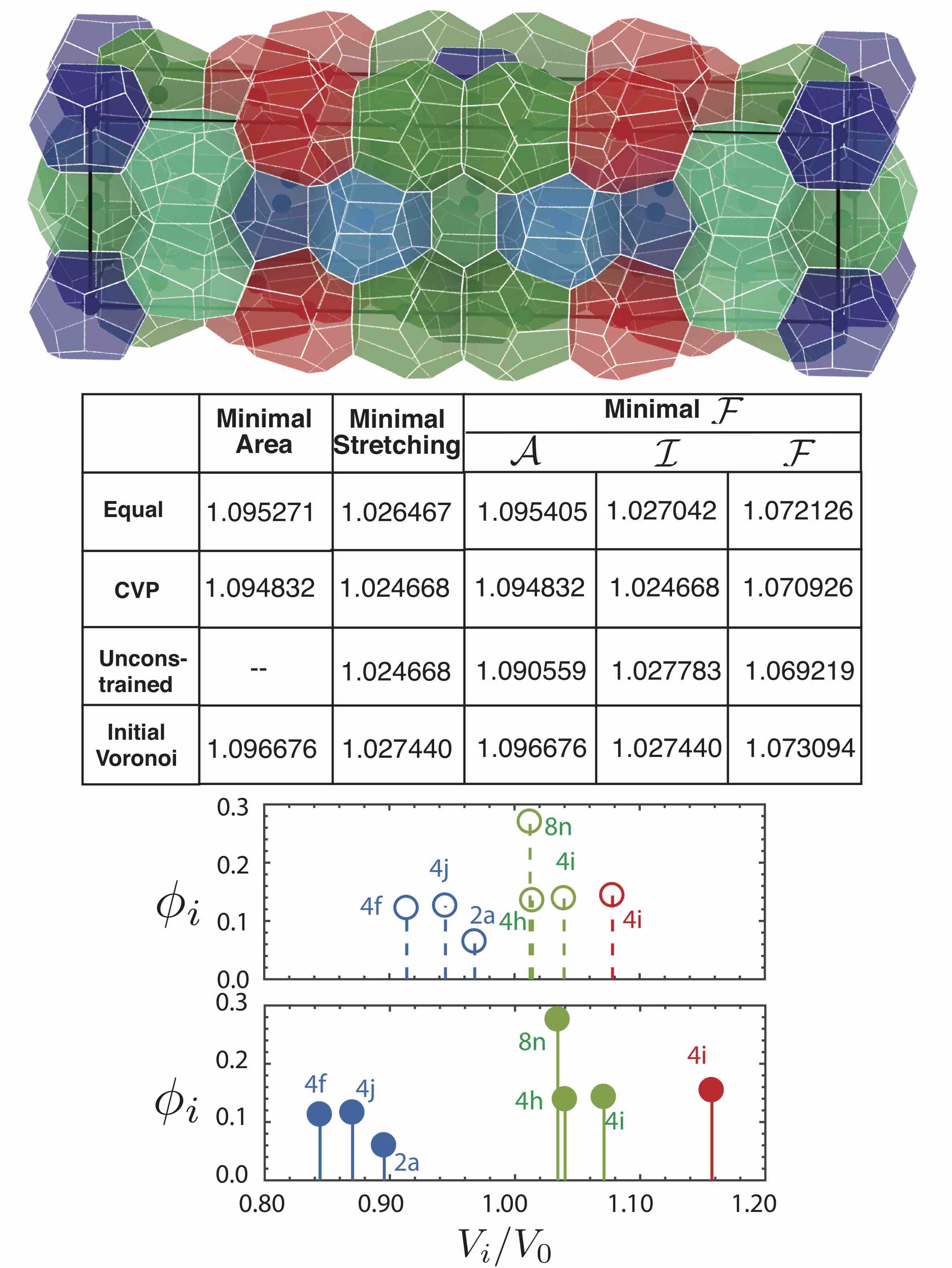}
\caption{$H$; space group: $Cmmm$; periodic cell:  (orthorhombic)  a:b:c=1:3.88:1; $n_{X}$ = 30; $\langle Z \rangle$ = 13.466; init. coords.: ref. ~\cite{H} (note that reference indicates Z12 spheres are situated at Wyckoff positions 2(a) and 4(e) whereas we have used positions 2(a) and 4(f))}
\end{figure}

\begin{figure}[H]
\center
\includegraphics[width=0.675\textwidth]{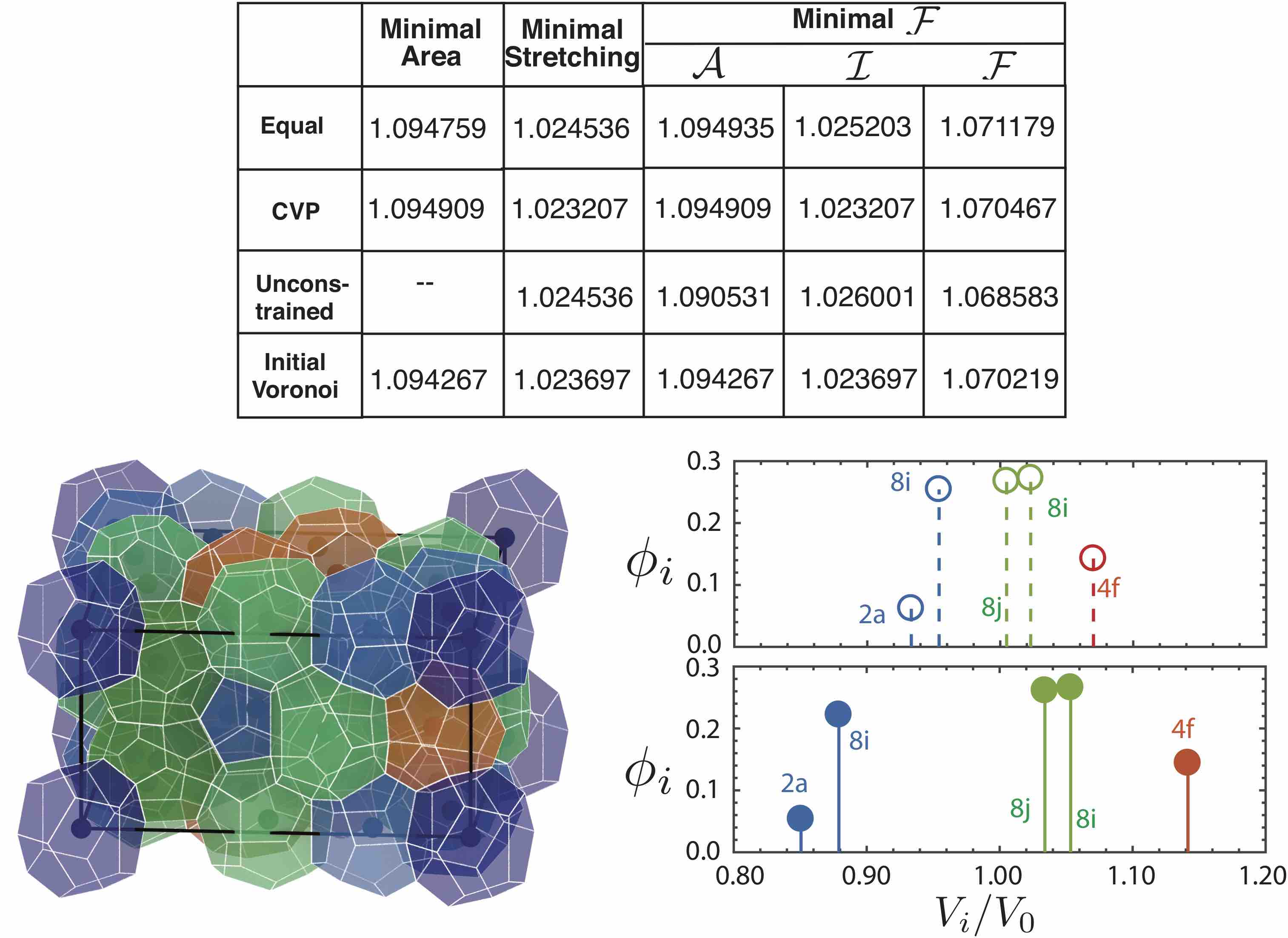}
\caption{$\sigma$; space group:  $P4_{2}/mnm$; periodic cell:  (tetragonal) a:b:c=1.9:1.9:1; $n_{X}$ = 30; $\langle Z \rangle$ = 13.466; init. coords.: ref.~\cite{Sigma}}
\end{figure}

\begin{figure}[H]
\center
\includegraphics[width=0.575\textwidth]{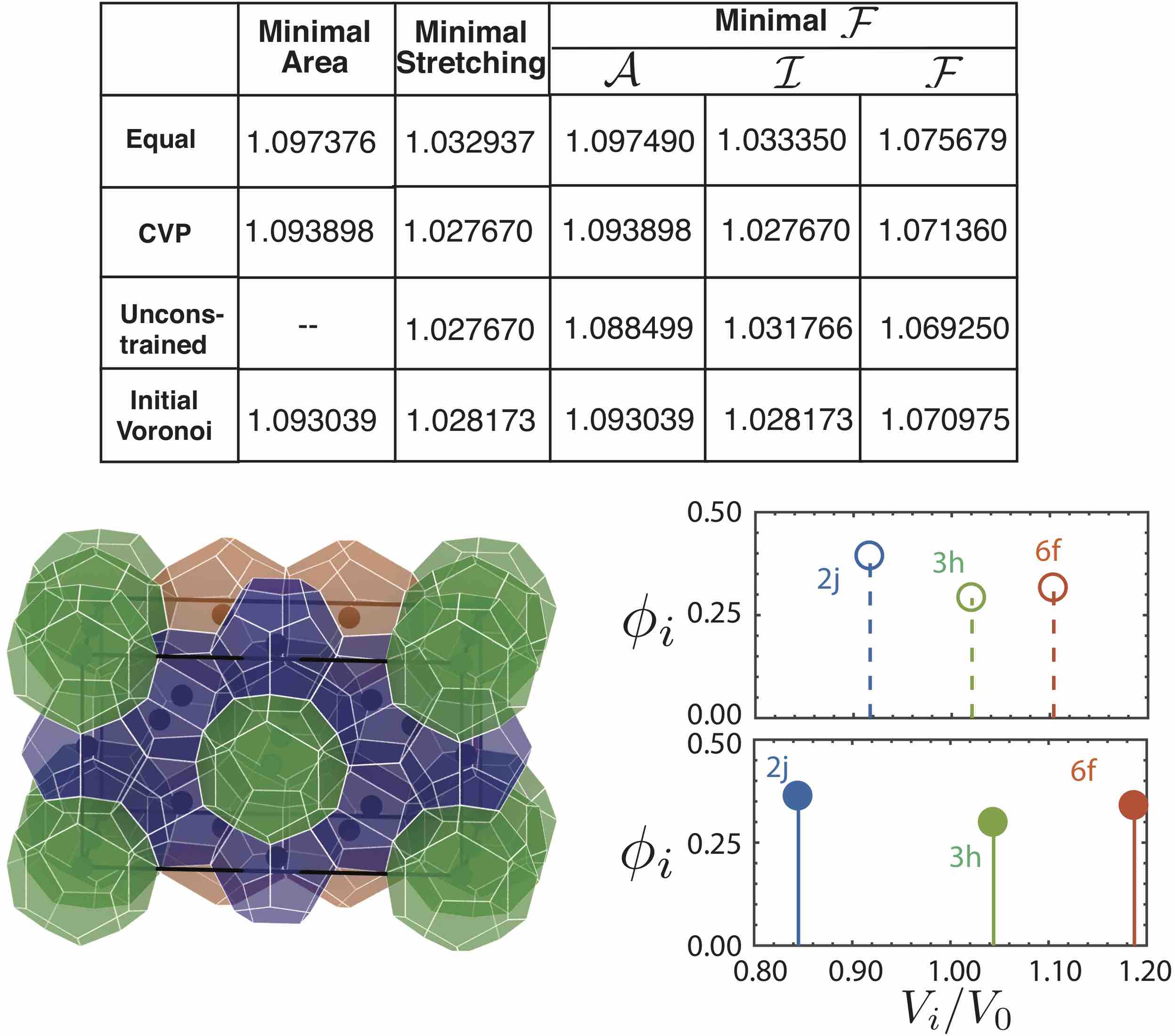}
\caption{$Z$; space group:  $P6/mmm$; periodic cell: (hexagonal*) a:b:c=1:1:0.993; $n_{X}$ = 7; $\langle Z \rangle$ = 13.428; init. coords.: ref.~\cite{Z} (*For SE calculations, we have used an equivalent orthorhombic unit cell with twice the number of cells than that of hexagonal unit cell)}
\end{figure}

\begin{figure}[H]
\center
\includegraphics[width=0.65\textwidth]{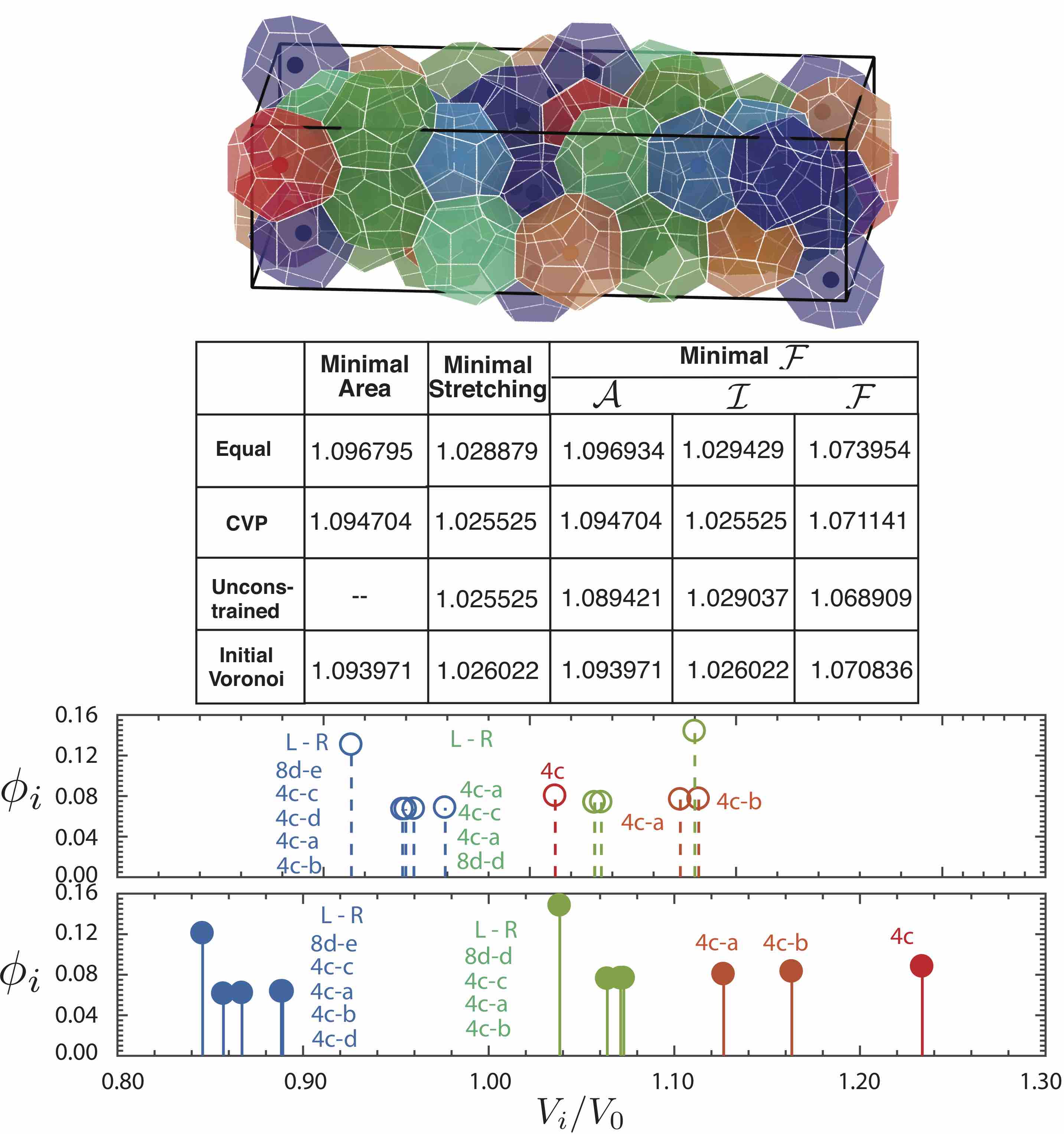}
\caption{$P$; space group: $Pbnm$, periodic cell: (orthorhombic) a:b:c=1.91:3.57:1; $n_{X}$ = 56; $\langle Z \rangle$ = 13.428; init. coords.: ref.~\cite{P}}
\end{figure}

\begin{figure}[H]
\center
\includegraphics[width=0.725\textwidth]{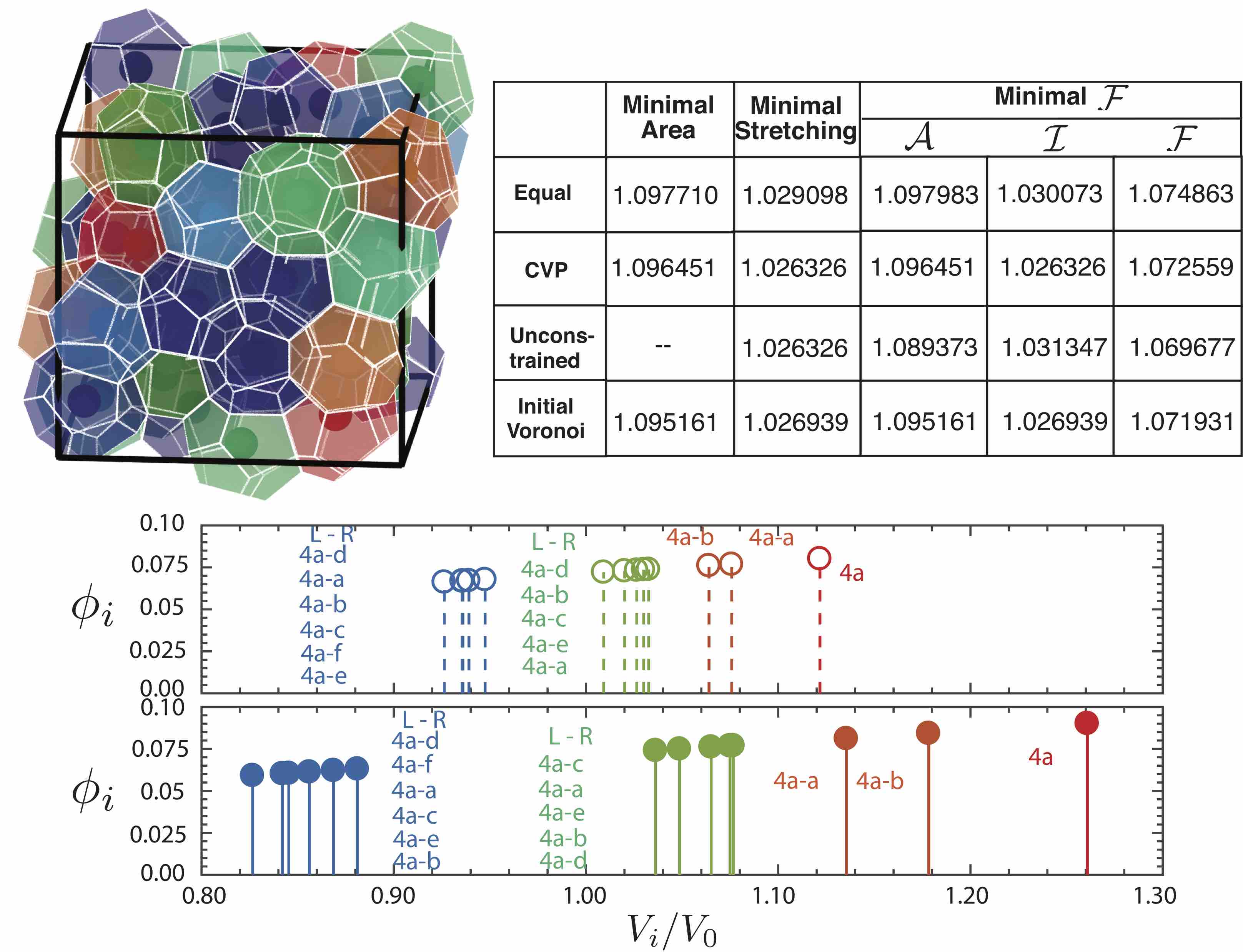}
\caption{$\delta$; space proup: $P2_{1}2_{1}2_{1}$; periodic cell: (orthorhombic) a:b:c=1.03:1.03:1; $n_{X}$ = 56; $\langle Z \rangle$ = 13.428; init. coords.: ref.~\cite{Delta}}
\end{figure}

\begin{figure}[H]
\center
\includegraphics[width=0.685\textwidth]{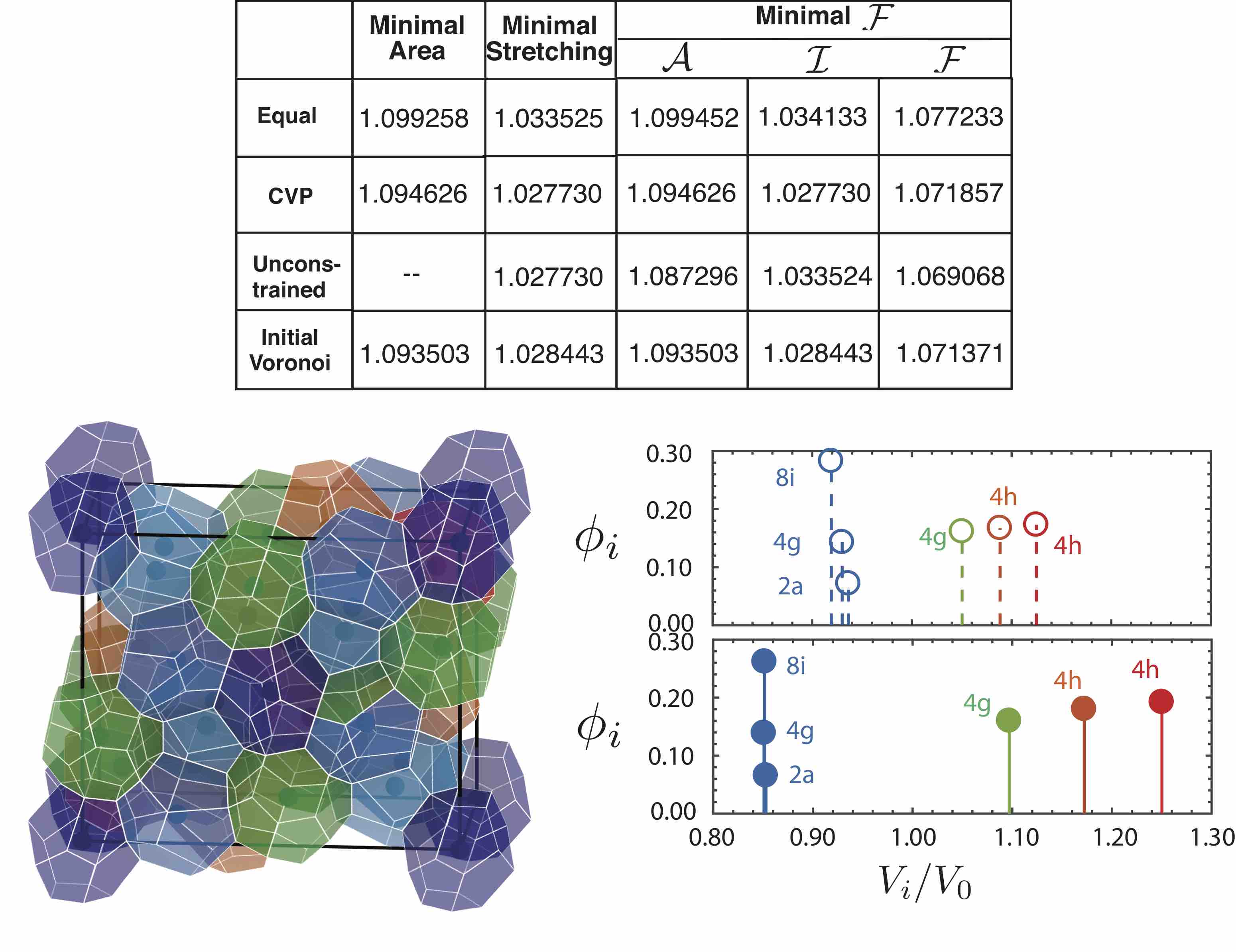}
\caption{$p\sigma$; space group: $Pbam$; periodic cell: (orthorhombic) a:b:c=1.95:1.64:1; $n_{X}$ = 26; $\langle Z \rangle$ = 13.385; init. coords.: ref.~\cite{Psigma}}
\end{figure}

\begin{figure}[H]
\center
\includegraphics[width=0.66\textwidth]{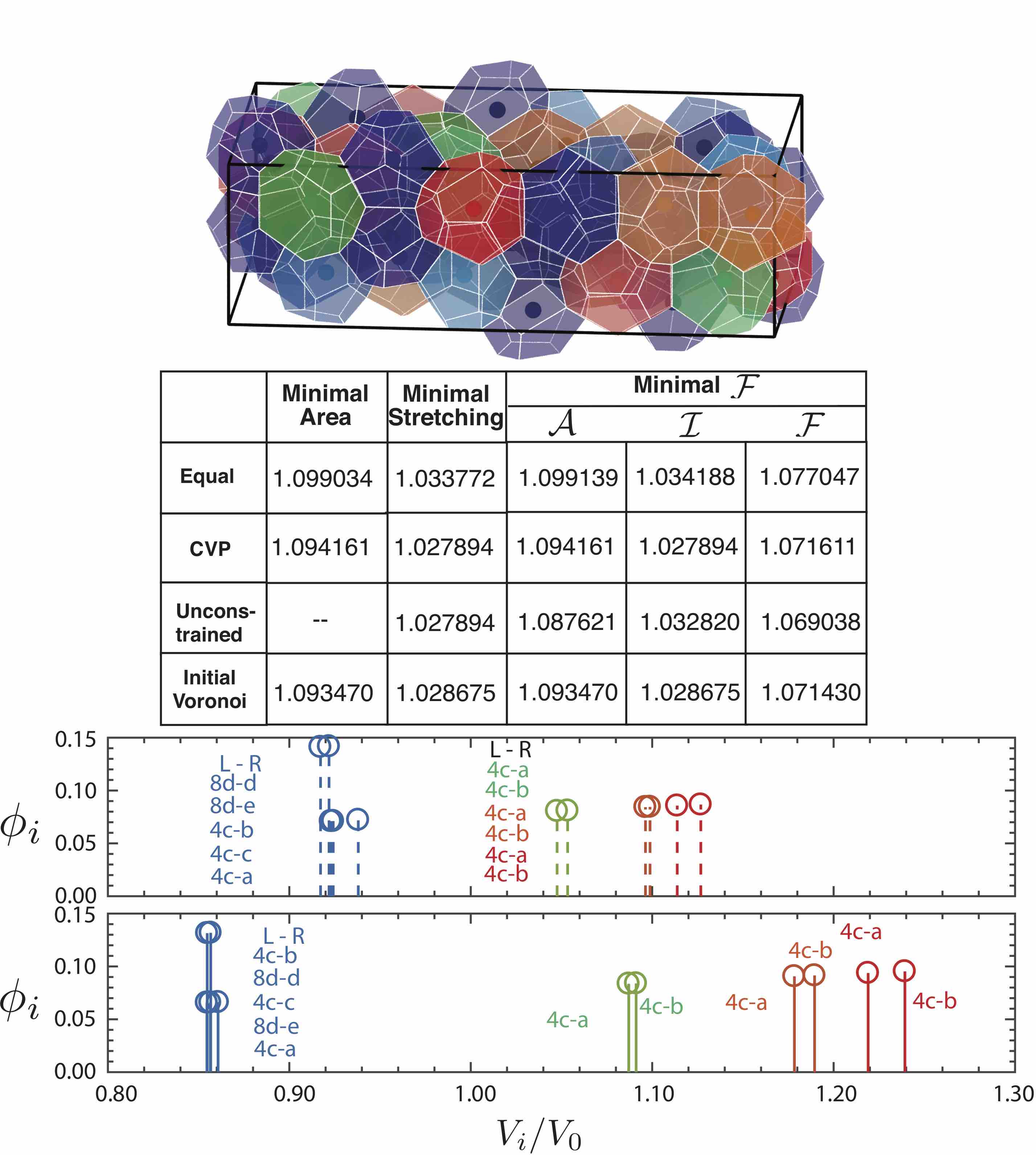}
\caption{$M$; space group: $Pnam$; periodic cell: (orthorhombic) a:b:c=1.89:3.3:1; $n_{X}$ = 52; $\langle Z \rangle$ = 13.385; init. coords.: ref.~\cite{M}}
\end{figure}

\begin{figure}[H]
\center
\includegraphics[width=0.67\textwidth]{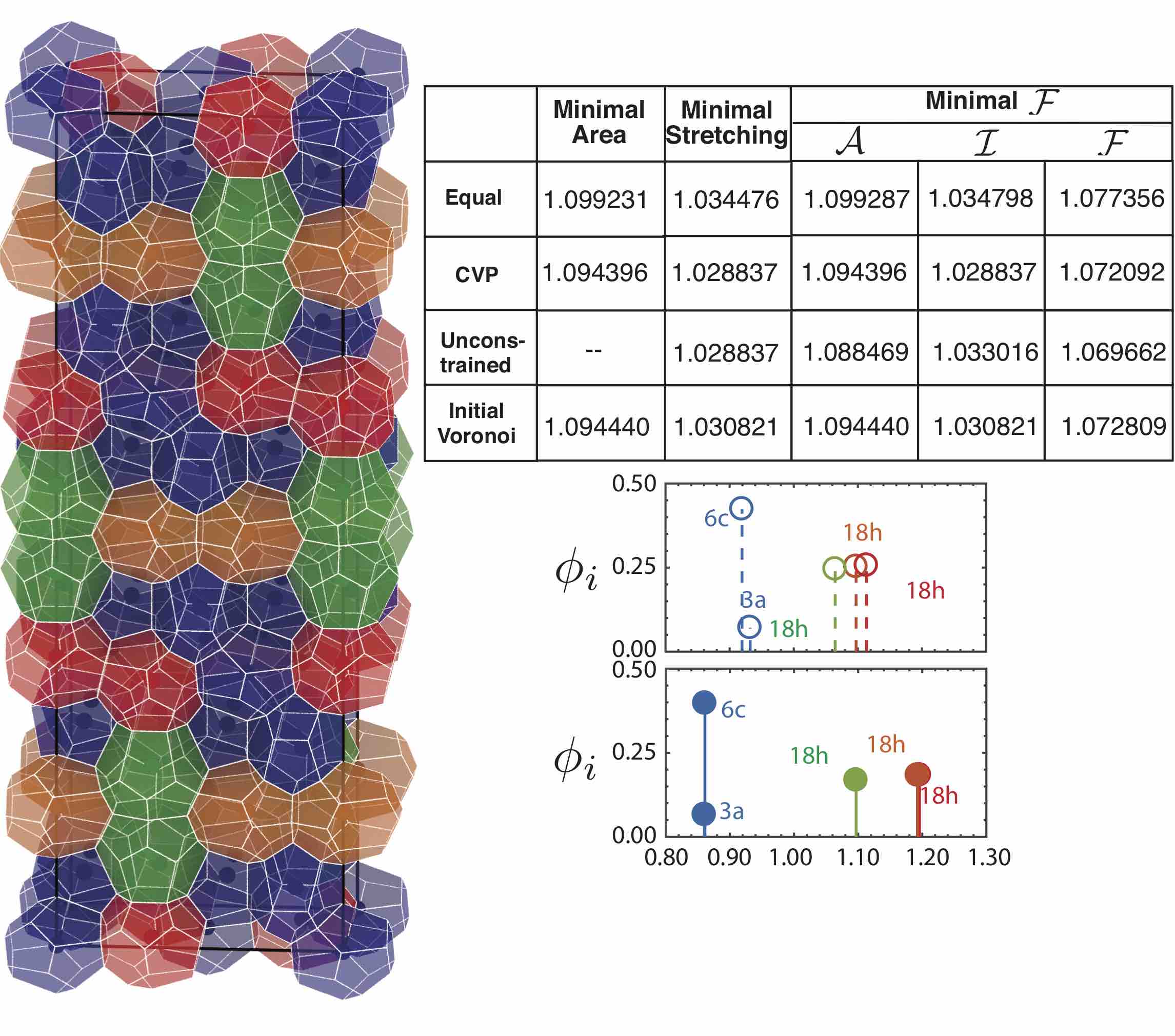}
\caption{$\mu$; space group: $R\bar{3}m$; periodic cell: (hexagonal*) a:b:c=1:1:5.2; $n_{X}$ = 39; $\langle Z \rangle$ = 13.385; init. coords.: $W_{6}Fe_{7}$ from ref.~\cite{A15C14C15Mu} (*For SE calculations, we have used an equivalent orthorhombic unit cell with twice the number of cells than that of hexagonal unit cell)}
\end{figure}

\begin{figure}[H]
\center
\includegraphics[width=0.7\textwidth]{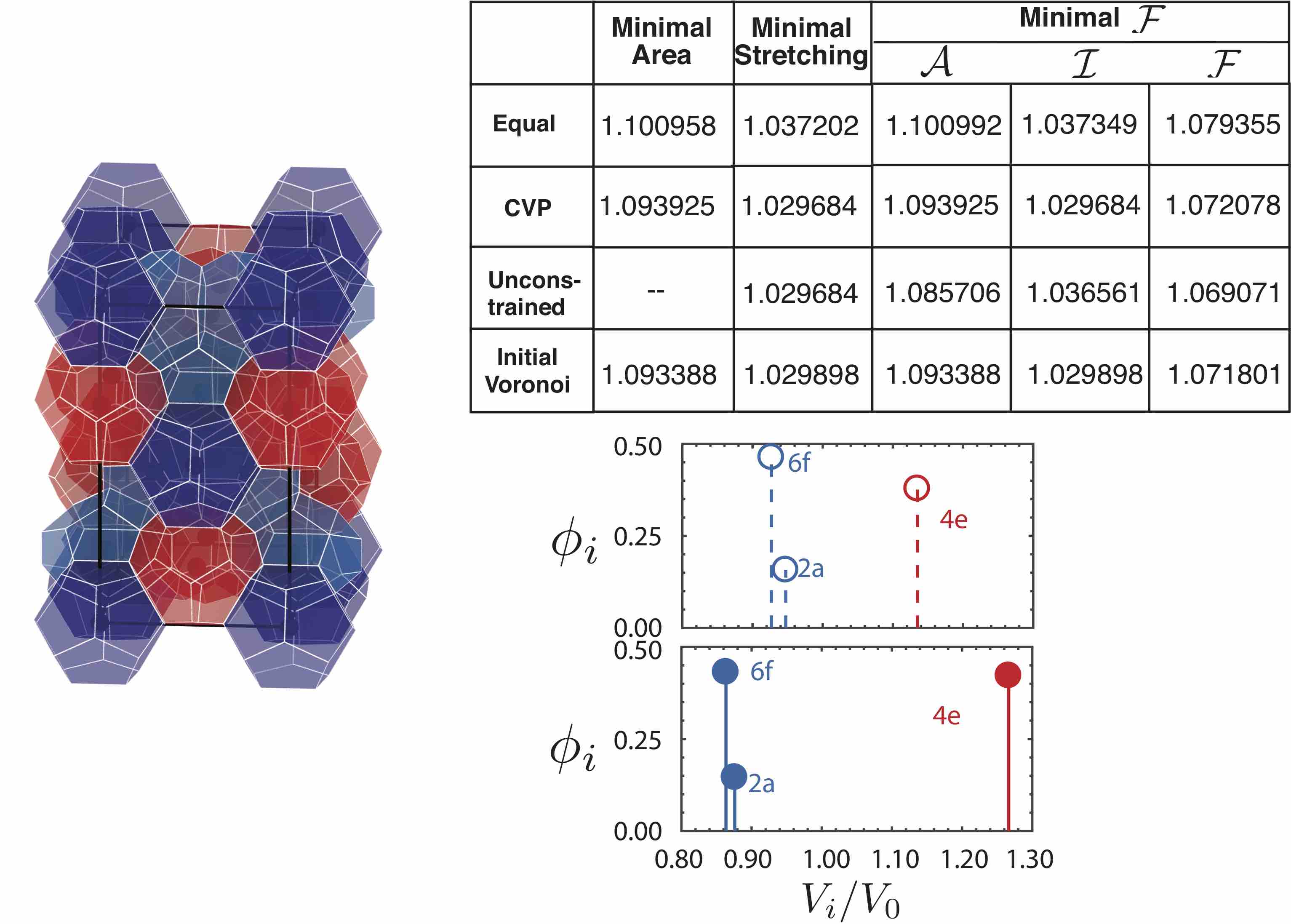}
\caption{$C14$; space group: $P6_{3}/mmc$; periodic cell: (hexagonal*) a:b:c=1:1:1.63; $n_{X}$ = 12; $\langle Z \rangle$ = 13.333; init. coords.: $MgZn_{2}$ from ref.~\cite{A15C14C15Mu} (*We have used an equivalent orthorhombic unit cell with twice the number of cells than that of hexagonal unit cell)}
\end{figure}

\begin{figure}[H]
\center
\includegraphics[width=0.725\textwidth]{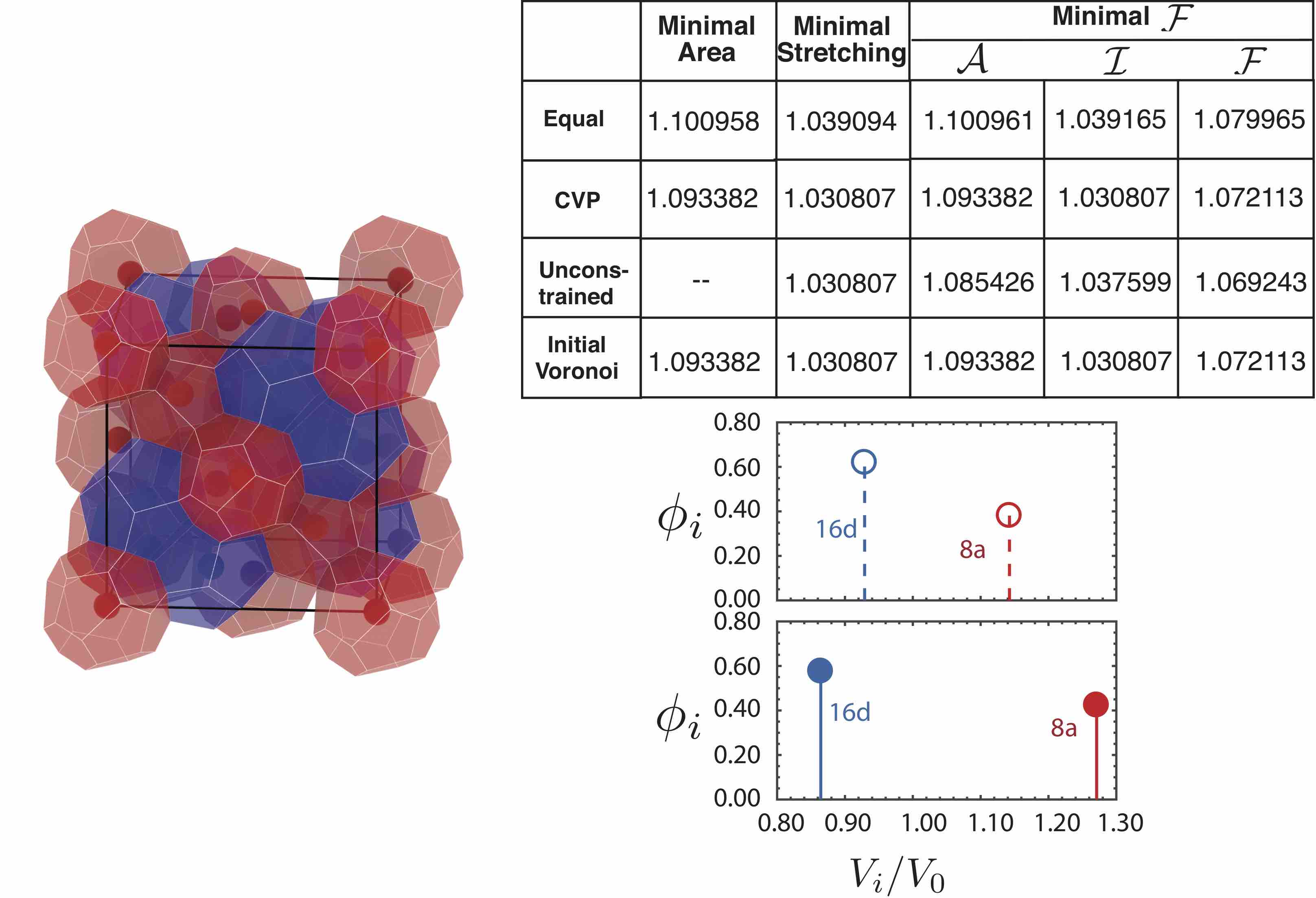}
\caption{$C15$; space group: $Fd\bar{3}m$, periodic cell: (cubic) a:b:c=1:1:1; $n_{X}$ = 24; $\langle Z \rangle$ = 13.333; init. coords.: $Cu _{2}Mg$ from ref. ~\cite{A15C14C15Mu}}
\end{figure}

\section{Self-consistent field theory of conformationally asymmetric diblocks}

We use self-consistent field theory (SCFT) of a Gaussian chain model of diblock copolymer melts~\cite{Matsen02} to predict structure and thermodynamics of a multi-chain qSD formation.  In particular, we consider a model where chains possess $N_{A}=fN$ and $N_{B}=(1-f)N$ segments of A and B type monomers each having statistical segment lengths as $a_{A}$ and $a_{B}= \epsilon^{-1} a_{A}$ respectively but having the same segment volume $\rho _{0} ^{-1}$, with the Flory-Huggins interaction parameter $\chi$ describing the enthalpic repulsion between A and B blocks.  In SCFT, the key statistical quantities are the chain distribution functions $q({\bf x} ,n)$ and $q^{\dagger}( {\bf x} ,n)$ which capture the statistical weights (constrained partial partition functions) of chains ``diffusing'' from their respective A and B ends to the $n$th segment located at position ${\bf x}$.  Following methods described in ref. \cite{Arora16} and elsewhere, these are determined self-consistently according to inter-segment interactions deriving from the mean compositions profiles $\phi _{A,B} ({\bf x}) = \frac{V}{N \mathcal{Q}} \int _{A,B}dn ~ q({\bf x},n) q^{\dagger}({\bf x},n) $, where $ \mathcal{Q} =  \int d^{\bf x} q({\bf x} ,n) q^{\dagger}( {\bf x} ,n)$ is the single chain partition function and $\int _{A,B}dn$ corresponds to the integration over the A or B block segments.

\subsection{Thermodynamics of SD phases}
\begin{figure}[H]
\center
\includegraphics[width=0.55\textwidth]{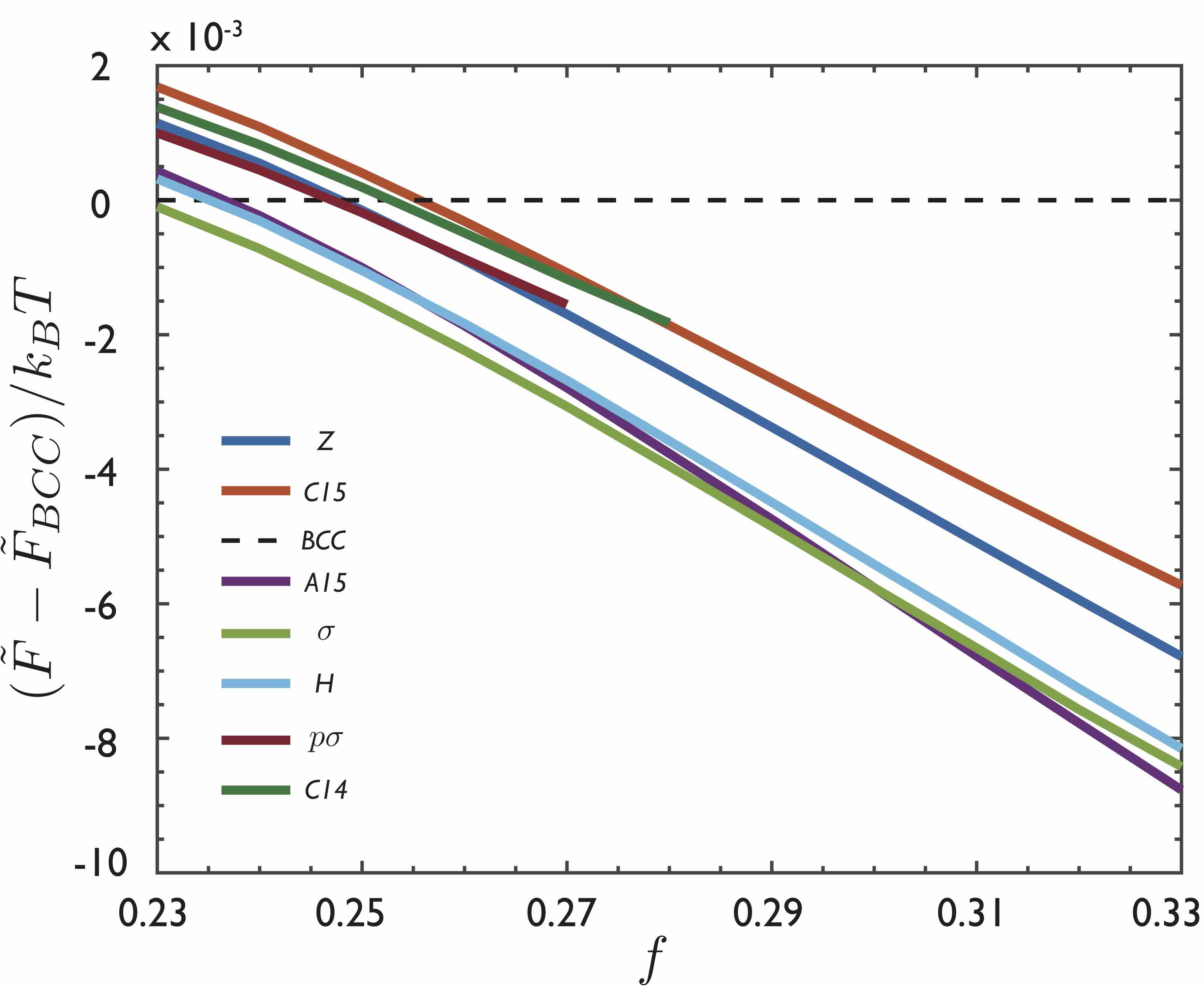}
\caption{\label{fig:e25} Relative Free energy per chain of FK lattices w.r.t BCC as a function of volume fraction for $\chi$N=25 }
\end{figure}

Here, we summarize results for thermodynamics of qSD phases, in comparison to DFM predictions, for conformational asymmetries $\epsilon > 1$.   For modest conformational asymmetry, i.e $\epsilon$=1.5 and 2, Xie {\it et al.}~\cite{ACShi14} and Kim  {\it et al.}~\cite{Kim17_1} have shown that $\sigma$ is the equilibrium for AB diblock copolymers in melt over a range of compositions between a stable BCC (low-$f$) and hexagonally ordered cylinders (high-$f$). Kim ~{\it et al.} have additionally reported results for FK candidates, $\sigma$, A15, Z, C14 and C15, for $\chi$N = 40, which we analyze in more detail. In Fig.\ref{fig:e25} we also report two new metastable structures H and $p\sigma$ for AB diblocks at $\chi$N = 25.  Although metastable, these phases all beat BCC over a range of $f$, and H is shown to be competitive with $\sigma$ and A15 over the entire range of metastable compositions studied, $0.23\leq f \leq0.33$.

\begin{figure}[H]
\center
\includegraphics[width=1\textwidth]{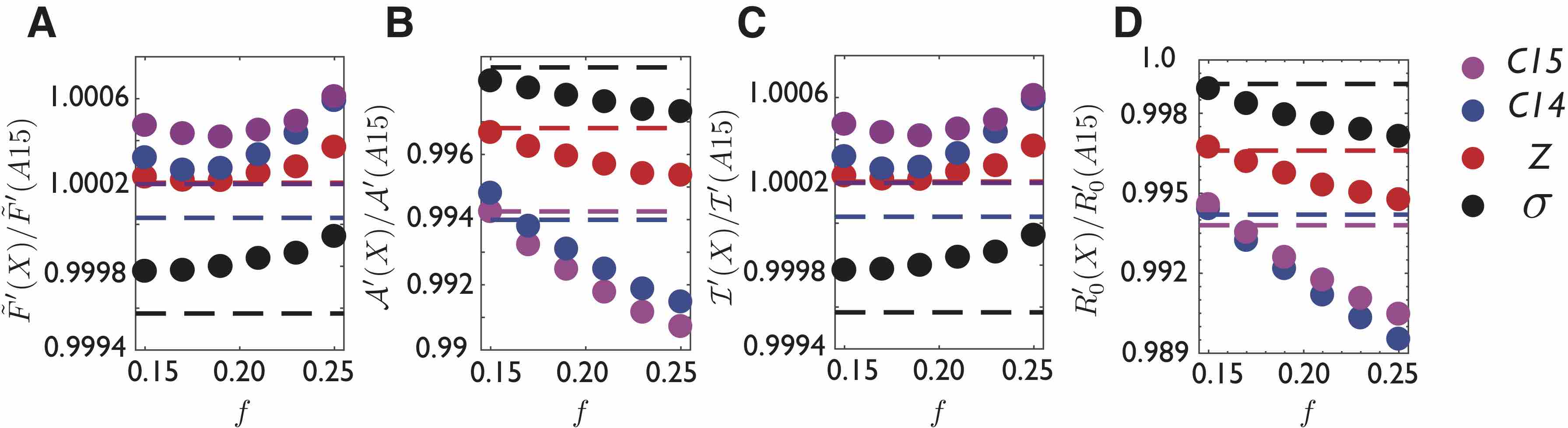}
\caption{\label{fig:scf_dfm}   (A) Relative free energy (B) Scaled enthalpy (C) Scaled entropy of FK lattices relative to A15 from SCF (circles) and DFM (dashed lines) (D) Relative mean domain sizes of FK lattices (relative to A15).  SCF results are for $\chi N = 40$ and $\epsilon = 2$. }
\end{figure}

To compare the DFM predictions to SCFT results, for $\sigma$, A15, Z, C14 and C15, at the highest segregation strength computed ($\chi N =40$) we normalize the free energy per chain by the value of A15, as plotted in Fig. \ref{fig:scf_dfm}A, since according to DFM, the free energy per chain for each structure is proportional to the same quantities (a function of $\chi N$, $\epsilon$ and $f$) that vary with cell geometry.  While DFM models are strictly constant with $f$ and SCFT results show at least slight variation of relative free energy with $f$, we note that the relatively close free energies of SCFT are remarkably consistent with scale of separation predicted by DFM predictions.

We also compare the relative ranking of $\sigma$, A15, Z, C14 and C15 in terms of the enthalpic and entropic contributions to the free energy per chain, $\tilde{F}'_{enthalpy}=V^{-1} \int d^3{\bf x} \chi\phi_A({\bf x}) \phi_B({\bf x} )$, and $\tilde{F}'_{entropy} =\tilde{F}'_{tot} - \tilde{F}'_{enthalpy}$, which are computed from SCFT solutions as described in ref. \cite{Matsen02} and elsewhere (here, primed quantities refer to values derived from SCF and unprimed quantities refer to their values from DFM). To extract strictly the geometric dependence of these thermodynamic quantities, we note from the DFM model (predicated on the strong-segregation and the polyhedral interface limits) that 
\begin{equation}
\tilde{F}_{enthalpy}=\frac{\gamma \cal A}{R_0};  \  \tilde{F}_{entropy}= \frac{\kappa}{2} {\cal I} R_0^2 \ \ \ \ \ ({\rm DFM})
\end{equation}
which motivates the definition of scaled-enthalpy $ {\cal A}'$ and scaled-entropy $ {\cal I}'$ computed from SCF results for $\tilde{F}_{enthalpy}'$ and $\tilde{F}_{entropy}'$, appropriately scaled by the mean sphere radius $R_{0}'$ (the radius of a sphere of equal mean volume to equilibrium SD for a given structure) according to
\begin{equation}
{\cal A}' \equiv\gamma^{-1}  \tilde{F}_{enthalpy}' R_{0}'; \ {\cal I}' \equiv \frac{2 \tilde{F}_{entropy}'} { \kappa R_{0}^{'2}}  \ \ \ \ \ ({\rm SCFT}) .
\end{equation}
This definition scales out the variation of enthalpic and entropic contributions due to the difference in mean domain sizes from structure to structure.  Fig. \ref{fig:scf_dfm} B and C plots the respective SCFT results for scaled enthalpy (${\cal A}' (X)$) and scaled entropy (${\cal I}' (X)$) normalized by the value for A15, and compared to DFM predictions.  Additionally, Fig. \ref{fig:scf_dfm} D plots the mean domain sizes $R'_0 (X)$ (relative to A15) computed from SCFT, which largely confirms that generic prediction of DFM, in eq. (\ref{eq: R0}) that structures corresponding to relatively small stretching costs favor relatively larger domain sizes (aggregation numbers per sphere).

\subsection{Geometric analysis of spherical domains}
\label{GeometrySCFTdetails}
\begin{figure}[H]
\center
\includegraphics[width=0.85\textwidth]{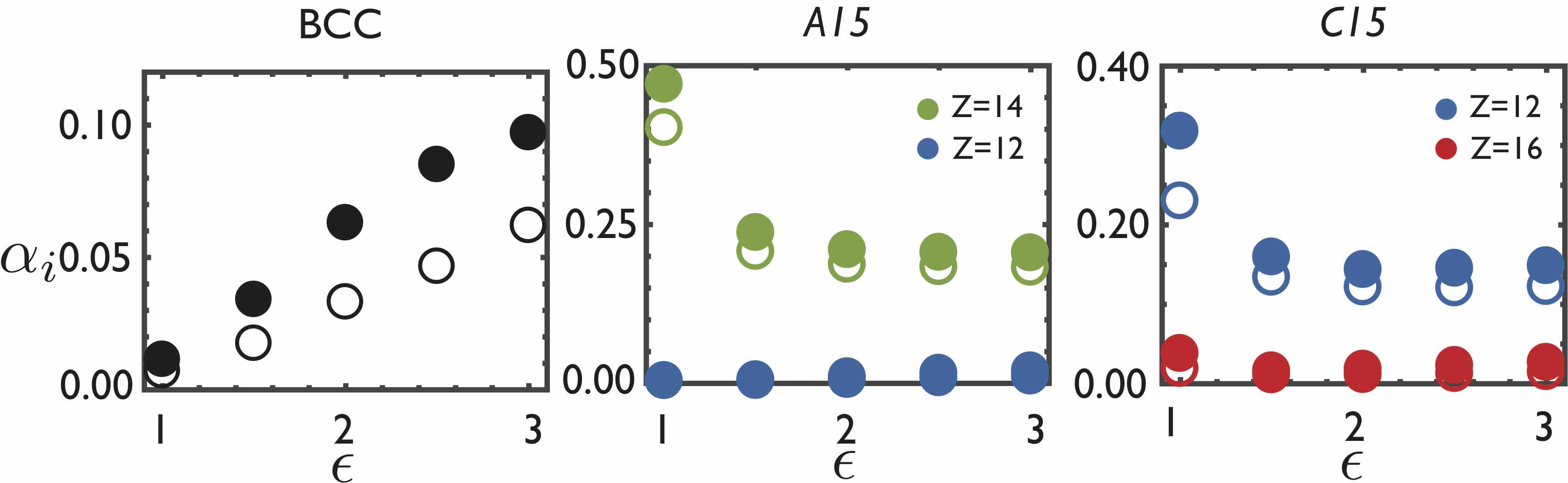}
\caption{\label{fig:alpha} Measure of areal distortion of AB interface of distinct domains for competing FK phases computed from SCF predictions at $\chi N =40$,  $f=0.29$ (open circles) and  $f=0.34$ (filled circles). }
\end{figure}

The geometry of SD core shapes predicted by SCFT are analyzed in terms of the AB interface, which can be extracted from the equilibrium composition profiles, specifically the isosurfaces where A and B have equal volume fractions, $\phi_A ({\bf x}) = \phi_B ({\bf x}) =0.5$.  From the isosurfaces, numerically extracted using MatLab, the total areas and enclosed volumes within each SD in the predicted SCFT structure can be directly computed.  Because the core blocks constitute a fixed fraction $f$ of the entire chain, the core volume accounts for the same fixed fraction of the entire qSD.  As shown in Fig.\ref{fig:3} for A15 and BCC at $f=0.29$, the areal distortion parameter, $\alpha_i$, of the core interface varies with conformational asymmetry.   In Fig. \ref{fig:alpha} we also show results at higher core composition, $f=0.34$.  These indicate that degree of polyhedral warping of interface increases with $f$, due to the enhanced proximity of the interface to the outer boundary (or cell ``wall'') between neighbor domains.  Fig. \ref{fig:alpha} also shows the variation of  $\alpha_i$ with $f$ for C15, highlighting the presence of two populations of SD in the structure: relatively spherical Z16 domains (low $\alpha_i$), and less spherical Z12 domains (higher $\alpha_i$).  

\begin{figure}[H]
\center
\includegraphics[width=0.95\textwidth]{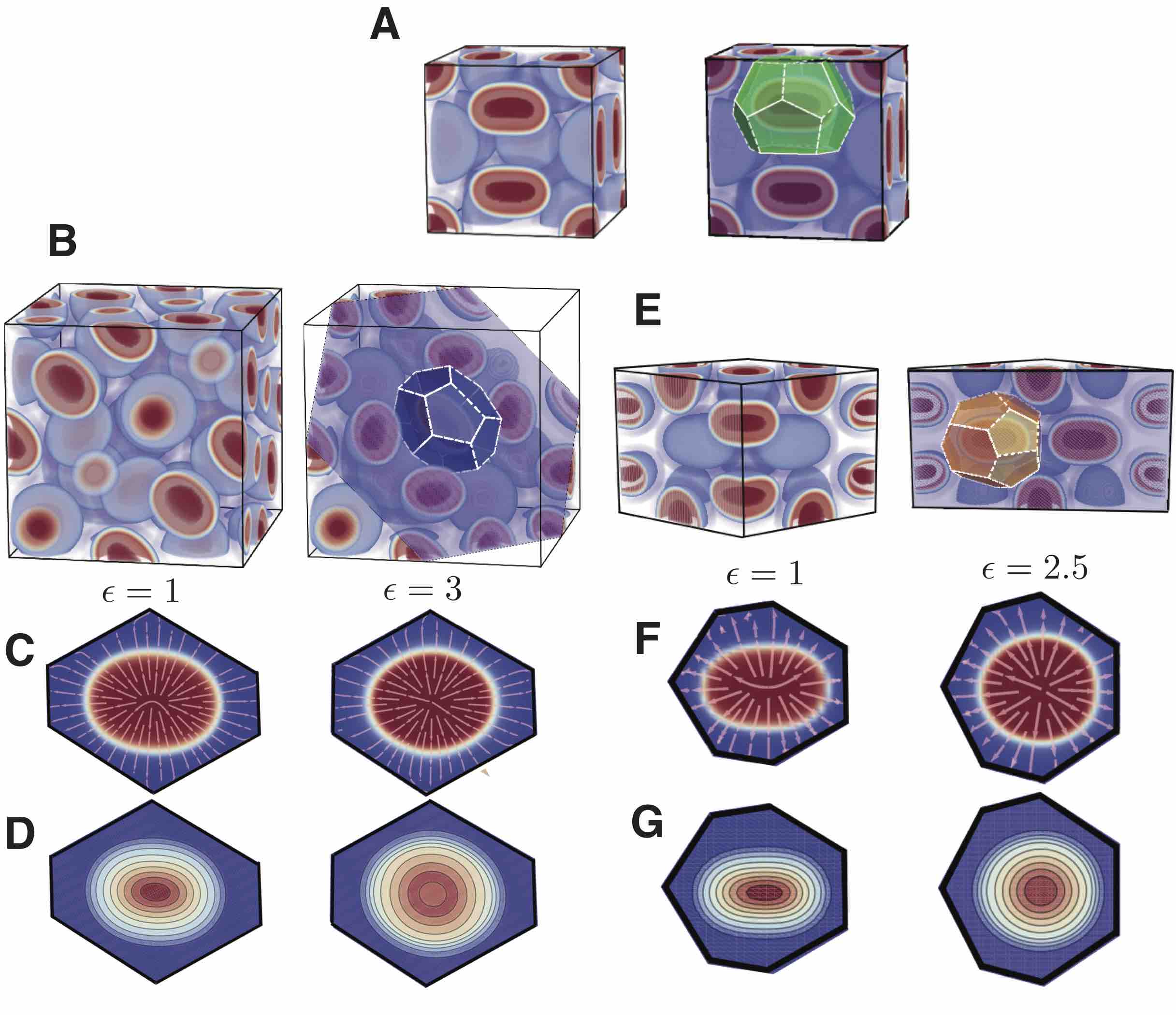}
\caption{\label{fig:sections} (A) 3D density plot of core block forming A15 from SCFT data at $f$=0.29, $\epsilon$=1, $\chi N = 40$ in primitive cell (on left) with the Z14 cell surrounding a corresponding qSD on the $[100]$ face shown in green (on right) corresponding to the sections shown in Fig.\ref{fig:3}.  (B) shows the same but for the C15 structure, and a 2D section through a $<111>$ plane through the center of a Z12 cell shown in blue (on right).  The composition and segment orientation for the Z12 domain of C15 are shown in (C), with the end distribution of the core A-block shown in (D) for conformationally symmetric and asymmetric cases.  (E) shows a hexagonal cell of Z phase from SCFT results at the same conditions at (A) and (B) (on left), with a cut through the center of the Z15 cell shown in orange (on right).  The composition and segment orientation for the Z15 domain of Z are shown in (F), with the end distribution of the core A-block shown in (G) for conformationally symmetric and asymmetric cases.}
\end{figure}

Like the case for the Z14 cells of A15 (shown in Fig.\ref{fig:3}), these higher-$\alpha_i$ cells of C15 and Z also undergo a discoidal-to-radial transition as $\epsilon$ is varied from 1 (conformationally symmetric) to $\approx 2$ (conformationally asymmetric).   To analyze the intra-domain structure of chain packing in more detail, we compute the {\it polar} orientational order parameter, $ {\bf t}_A ({\bf x})$, of A block segments using methods described in ref. \cite{Prasad17},
\begin{equation}
 {\bf t}_A ({\bf x})= \frac{V}{6 N \mathcal{Q}} \int _{A} dn [q \nabla q^{\dagger} - q^{\dagger} \nabla q] ,
\end{equation}
where the vector orientation of the segments is defined to point from the free A end towards the junction point along the chains.  These orientational profiles are shown in 2D sections through spherical domains of BCC and A15 in Fig.\ref{fig:3}.  In Fig. \ref{fig:sections}  we also show 2D cuts through Z12 domains and Z15 domains of C15 (B,C) and Z (E,F), respectively.  Note that only streamlines of $ {\bf t}_A ({\bf x})$ are shown in these figure, and thus only the local orientation of A segments, but not the degree of alignment, is visible.  Both of these domains show a trend consistent with the sub-domain morphology of the Z14 cell in A15.  For conformationally symmetric chains ($\epsilon =1$), the interface shape is more oblate than the polyhedral cell enclosing the domain (which itself can be observed from the flow lines about the orientational order parameter), and the core regions are composed of a quasi-lamellar ``puck'' encircled by quasi-toroidal rim.  In contrast, when conformational asymmetry imposes a sufficiently larger cost on coronal stretching ($\epsilon \gtrsim 2$), the core interface shape is more consistent with an affinely shrunk (and somewhat rounded) copy of the polyhedral cell, and segment orientation becomes consistent with uniformly radial extension of the chains from the domain center.   The sub-domain distinctions between radial and discoidal domain shapes is further highlighted by comparing the respectively focussed vs. spatially spread distributions of A-block ends, computed from $q^\dagger ({\bf x}, n=0)$ and shown in Fig. \ref{fig:sections} D and G.  

\end{document}